\documentclass{elsarticle} %review
\usepackage[english]{babel}

\usepackage{amssymb, mathtools, amsthm, mathrsfs}
\usepackage{subcaption}
\usepackage{graphicx}
\graphicspath{{_pics/}}

\usepackage{fullpage}
\usepackage{comment}

\usepackage{xcolor}

\usepackage{hyperref}

\usepackage{tikz}
\usetikzlibrary{arrows, math, decorations.pathreplacing, calc}

\tikzset{
    >=latex',
    interface/.style={blue, very thick},
    boundary/.style={interface, postaction={draw, decorate, thick,
        decoration={border, angle=#1, amplitude=0.3cm, segment length=2mm, pre length=1mm}}},
    ghost/.style={fill=gray!50},
    eqref/.style={red},
    mapping/.style={eqref, dashed},
    bolddot/.style={shape=circle, fill=black, scale=0.5},
}

\journal{Journal of \LaTeX\ Templates}

\newcommand{\dd}{\mathrm{d}}
\newcommand{\pder}[2][]{\frac{\partial#1}{\partial#2}}

\newcommand{\pderder}[3][]{\frac{\partial^2#1}{\partial#2\partial#3}}
\newcommand{\pderderder}[4][]{\frac{\partial^3#1}{\partial#2\partial#3\partial#4}}
\newcommand{\Pder}[2][]{\partial#1/\partial#2}

\newcommand{\Set}[2]{\{\,{#1}:{#2}\,\}}
\newcommand{\OO}[1]{O(#1)}
\newcommand{\transpose}[1]{#1^\mathsf{T}}
\DeclarePairedDelimiter\autobracket()       % mathtools are needed
\newcommand{\br}[1]{\autobracket*{#1}}

\DeclareMathOperator{\sgn}{sgn}

\newcommand{\dxi}{\dd{\boldsymbol{\xi}}}
\newcommand{\bxi}{\boldsymbol{\xi}}
\newcommand{\bpsi}{\boldsymbol{\psi}}
\newcommand{\bv}{\boldsymbol{v}}
\newcommand{\bq}{\boldsymbol{q}}
\newcommand{\bn}{\boldsymbol{n}}
\newcommand{\be}{\boldsymbol{e}}
\newcommand{\bm}{\boldsymbol{m}}
\newcommand{\bT}{\boldsymbol{T}}
\newcommand{\bdot}{\boldsymbol{\cdot}}

\newcommand{\bx}{\boldsymbol{x}}
\newcommand{\equil}[1]{#1^\mathrm{(eq)}}
\newcommand{\hermite}[1]{#1^\mathrm{(H)}}
\newcommand{\NSF}[1]{#1^\mathrm{(NSF)}}
\newcommand{\refer}[1]{#1_0}
\newcommand{\LB}{\mathrm{LB}}
\newcommand{\DV}{\mathrm{DV}}
\newcommand{\Herm}{\mathcal{H}}
\newcommand{\cut}{{(\mathrm{cut})}}

\newcommand{\xiaj}{\xi_{\alpha,j}}
\newcommand{\xibj}{\xi_{\beta,j}}
\newcommand{\xigj}{\xi_{\gamma,j}}
\newcommand{\caj}{c_{\alpha,j}}
\newcommand{\cbj}{c_{\beta,j}}

\newcommand{\Aa}{a_{\alpha}}
\newcommand{\Aab}{a_{\alpha\beta}}
\newcommand{\Aabg}{a_{\alpha\beta\gamma}}
\newcommand{\Ha}{H_{\alpha}}
\newcommand{\Hab}{H_{\alpha\beta}}
\newcommand{\Habg}{H_{\alpha\beta\gamma}}

\begin{document}

\begin{frontmatter}

\title{Kinetic multiscale scheme based on the discrete-velocity and lattice-Boltzmann methods}

\author[ccas]{V.V.~Aristov}
\author[ccas]{O.V.~Ilyin}
\author[skoltech,ccas]{O.A.~Rogozin\corref{cor}}
\cortext[cor]{Corresponding author}
\ead{oleg.rogozin@phystech.edu}

\address[ccas]{Dorodnicyn Computing Center,
    Federal Research Center "Computer Science and Control" of Russian Academy of Science, Moscow, Russia}
\address[skoltech]{Center for Design, Manufacturing, and Materials,
    Skolkovo Institute of Science and Technology, Moscow, Russia}

\begin{abstract}

A novel hybrid computational method based on the discrete-velocity (DV) approximation,
including the lattice-Boltzmann (LB) technique, is proposed.
Numerical schemes for the kinetic equations are used in regions of rarefied flows,
and LB schemes are employed in continuum flow zones.
The schemes are written under the finite-volume (FV) formulation to achieve
the flexibility of local mesh refinement.
The truncated Hermite polynomial expansion is used for matching of DV and LB solutions.
Special attention is paid to preserving conservation properties in the coupling algorithm.
The test results obtained for the Couette flow of a rarefied gas
are in excellent agreement with the benchmark solutions,
mostly thanks to mesh refinement (both in the physical and velocity spaces) in the Knudsen layer.

%In the present paper, a novel hybrid fluid--kinetic computational approach based on the discrete-velocity (DV) approximation including lattice-Boltzmann (LB) technique is proposed. Numerical schemes for a kinetic equation are used in regions of rarefied flows and LB schemes are used in continuum flow zones. The problem devoted to matching solutions of DV and LB in buffer cells is solved using the expansion of the solutions of DV and LB models on Hermite polynomials and matching the expansion coefficients. A special attention is paid to the recent high-order LB models. Some variants of LBM are considered and the efficiency of the hybrid method is evaluated. The Couette-flow problem is investigated as a test. The criteria transformation from the continuum region to kinetic one is proposed and the influence of the point transformation is studied. The CPU times for the hybrid scheme and for BGK scheme are compared. The good correspondence with the benchmark solution is obtained: the computations on the basis of the hybrid method are close to the well-known tabulated solutions of the Couette flow.

\end{abstract}

\begin{keyword}
%kinetic--kinetic coupling,
%kinetic multiscale scheme,
hybrid numerical method,
discrete-velocity method,
lattice-Boltzmann method,
domain decomposition,
Knudsen layer.
\end{keyword}

\end{frontmatter}

% \linenumbers
\tableofcontents

\section{Introduction}\label{sec:intro}

%%% Effective numerical methods for multiscale problems
Thus far, effective numerical simulation of multiscale flows has remained a challenging problem
despite the efforts of many researchers.
This is due, in particular, to complicated flow structures, where small-scale highly nonequilibrium regions
coexist with large-scale equilibrium zones.
The use of the kinetic equation in all regions is very demanding from a computational point of view.
On the other hand, the fluid-dynamics models provide an efficient approximation of near-equilibrium flows,
but this kind of description is not adequate in regions where the velocity distribution function (VDF)
is far from the Maxwellian and the contribution of high moments cannot be neglected.

%%% Aristov's version:
% Therefore, computational fluid dynamics is being confronted with the challenge of constructing
% the unified numerical methods suitable for simulation of multiscale flows.
% The numerical solution of kinetic equations in stiff regimes represents a challenge in the construction of computational methods.
% one would like to avoid the expensive cost of solving the kinetic equation in regions well described by continuum fluid models,
% since the latter are easily solvable by classical numerical methods.

%%% Citation from [Meng, Reese, Zhang 2014]:
% Therefore, to strike a balance between computational costs and simulation accuracy,
% multiscale schemes are being developed that take advantage of both kinetic and continuum-fluid solvers,
% i.e., deploying a kinetic solver only in the rarefied flow regions and a continuum solver in the hydrodynamic regions.

%%% Classic hybrid schemes
There are two main approaches on how to deal with the multiscale problems~\cite{Dimarco2014}.
The first one employs different kinds of representations for equilibrium and nonequilibrium parts of the solution
in the entire computational space, while the second one handles the problem by dividing the physical domain
into the highly rarefied and near-equilibrium regions using some criterion of domain decomposition.
The fluid--kinetic coupling is a natural and effective approach for the description of multiscale flows.
Coupling of the Boltzmann and Euler or Navier--Stokes (NS) equations
is a canonical example of such hybrid schemes~\cite{Bourgat1996, Tallec1997}.

%and has also been elaborated and applied within the Unified Flow Solver (UFS)~\cite{Kolobov2007}.
%%% Information from [Dimarco, Pareschi 2014]
%1. Asymptotic-preserving schemes
%2. Dynamic domain decomposition strategies
%    A. Dynamic fluid--kinetic coupling methods:
%        a) A moving interface method: a stationary smooth transition strategy was proposed for this coupling.
%        b) A micro–macro moving interface method (Degond et al. 2010): micro–macro decomposition of the VDF
%3. Hybrid methods
%    A. Low-variance deviational Monte Carlo methods
%    % Variance reduction methods are a popular way to improve the accuracy of Monte Carlo methods by reducing the amount of fluctuations in the results (Caflisch 1998).
%    B. Moment-guided Monte Carlo methods
%    % The basic idea described here consists in obtaining reduced-variance Monte Carlo methods by forcing the statistical samples to match prescribed sets of moments given by the solution of deterministic macroscopic fluid equations (Degond et al. 2011, Dimarco 2013).
%    C. Hybrid multiscale methods (hybrid representation of the solution)
%    % different kind of representation of the solution as an equilibrium and nonequilibrium part, first introduced in Pareschi and Caflisch (1999).
%    % the solution in each cell is represented as a combination of two different parts, a stochastic particle representation of the nonequilibrium fraction and a deterministic representation of the equilibrium part

%%% Classical kinetic schemes for fluid dynamics
The numerical schemes based on the kinetic description of the fluid and capable of reproducing the Euler and NS dynamics
have been suggested independently and widely used since the early 1980s~\cite{Potkin1975, Pullin1980, Reitz1981, Aristov1983}.
Later these pioneer kinetic schemes have been significantly developed
in~\cite{Elizarova1985, Deshpande1986, Prendergast1993, Chou1997, Ohwada2004Xu, Ohwada2004Kobayashi, Ohwada2006}.
The cellular-automata approximation for the NS equations was developed in the middle of the 80s~\cite{Frisch1986}.
Finally, the lattice-gas model based on the BGK equation was proposed at the beginning of the 90s~\cite{Qian1992}.
It gave rise to a broad class of numerical methods called lattice-Boltzmann (LB) methods~\cite{Higuera1989, Benzi1992, Succi2001}.

%%% Models used in the LB method
The LB models can be roughly divided into two classes: low-order and high-order.
The low-order LB models recover the correct hydrodynamics for small Mach numbers,
while the high-order models are able to reproduce full Navier--Stokes--Fourier equations
and, moreover, some rarefied-gas effects~\cite{Succi2001, Shan2006}.
The LB method is a flexible framework for developing sophisticated models.
Many recent works are devoted to extending the applicability of the LB method beyond the Navier--Stokes level.
In particular, models that correctly describe the Burnett-level dynamics is formulated in~\cite{Xu2018, Xu2019}.
Special-purpose methodologies are proposed to capture rarefied-gas effects in the highly nonequilibrium Knudsen layer
arising from the kinetic boundary conditions~\cite{Feuchter2016, Ambrus2014, Ambrus2016, Ambrus2016mixed}.
Another promising approach is based on the regularization procedure for LB models~\cite{Mont2015}.

%%% Relation to Broadwell-type DV models and LB-DV comparison
It is worth emphasizing that there is a fundamental relationship between the LB and DV methods.
For instance, one can cite a phrase from~\cite{Rivet2001}:
``This type of discrete kinetic theory can be seen as the ancestor of the lattice gas approach.''
The LB method is genetically related to the Broadwell-type models~\cite{Broadwell1964shock, Gatignol1975},
which use a small number of discrete velocities to reproduce some relevant features of the Boltzmann equation.
The DV method typically uses lattices with a large number of discrete velocities,
which can accurately approximate highly nonequilibrium VDF.
In contrast, discontinuities and sharp variations of the VDF cannot be reproduced
on a span of the truncated basis of smooth polynomials underlying the LB models.

%%% Known coupling schemes for LB
The idea of coupling the low-order and high-order LB models is presented in~\cite{Meng2011}.
The hybrid approach based on the DSMC and LB methods is proposed in~\cite{Staso2016long, Staso2018}.
Unlike DSMC, the deterministic methods for solving the Boltzmann equation do not produce statistical noise.
Therefore, hybrid approaches based on them appear to be more promising.
The possibility of merging the DV and LB methods is noticed in~\cite{Succi2016},
while the first results based on matching of the half-range fluxes have been presented in~\cite{Ilyin2018}.

%%% Some extra details on DV methods
% The DV schemes with the large number of discrete velocities are used in direct methods for solving BE, BGK, S-model or other kinetic equations.
% The DV method is applied with the combination of Monte Carlo or quasi-Monte Carlo procedures for evaluating collision integrals
% and for computing the appropriate moments which are used in the collision integrals of the model kinetic equations.
% For the near-equilibrium zones, the small number of discrete velocities can be considered.
% Therefore, one can expect that LB approaches are fit for describing flows in these regions.

%%% Our proposition
The proposed hybrid kinetic approach is based on coupling the DV and LB methods.
The DV method accurately describes nonequilibrium regions,
while the LB method provides an efficient approximation in the continuum regions.
To couple solutions between the DV and LB subdomains,
the VDF is projected onto the truncated Hermite basis in the buffer zone.
The kinetic breakdown criterion determines the position of the coupling interface.
The first implementation of this method is reported in~\cite{Aristov2019}.
In the present paper, the accuracy of the method has been significantly improved
by employing nonuniform grids for the DV approximation in the Knudsen layer.

%%% FV formulation
The classical LB methods enjoy their efficiency coming from the highly symmetric discrete physical space and time.
However, uniform Cartesian meshes lack flexibility and, therefore, local mesh refinement.
There are several approaches how to work around this limitation.
The LB method is easily extended for arbitrary unstructured meshes
under the FV formulation~\cite{Succi1992, Peng1999, Patil2009, Li2016}.
In the present paper, this strategy is adopted, specifically to refine mesh near the boundary.

%%% Emphasizing the difference with DiStaso method
Ideologically, the present work is similar to the works~\cite{Staso2016long, Staso2018},
but there is a significant difference.
The proposed hybrid approach is based on the unified DV approximation,
whereas the DSMC method has a fundamentally different representation of the VDF.
In this sense, coupling DV and LB models under the single FV formulation is more natural.

%%% Plan of the paper
The plan of the present paper is as follows.
The governing equations and nondimensional variables are introduced in Sec.~\ref{sec:equations}.
The DV approximation of the VDF underlying both DV and LB method is outlined in~\ref{sec:dv}.
The mapping method between DV and LB models is described in Sec.~\ref{sec:mapping}.
Details of the numerical schemes and algorithms are presented in Sec.~\ref{sec:numerics}.
Numerical solutions of the Couette-flow problem are obtained by the DV model,
various LB models, and their hybrid combinations.
They are illustrated and compared with each other for various Knudsen numbers in Sec.~\ref{sec:results}.
Possible equilibrium breakdown parameters and computational efficiency of the hybrid approach
is also analyzed in Sec.~\ref{sec:results}.
Perspectives of the kinetic multiscale methods based on coupling the DV and LB solutions
are discussed in Sec.~\ref{sec:summary}.
The peculiar details of the finite-volume scheme are included in~\ref{sec:limiter}.
Finally, a set of supplementary formulas for the benchmark solution of the linear Couette-flow problem,
which is used to validate the investigated numerical methods, is provided in~\ref{sec:benchmark_solution}.

\section{Main equations}\label{sec:equations}
%%%%%%%%%%%%%%%%%%%%%%%%%%%%%%%
%%% [Rogozin]: there should be nothing about Couette flow, discretization, methods, and algorithms in "Main equations"!
%%%%%%%%%%%%%%%%%%%%%%%%%%%%%%%

%%% Nondimensional variables
We first introduce the notation for describing a dilute gas.
Let \(L\), \(\refer\rho\), \(\refer{T}\), \(\refer{c} = \sqrt{R\refer{T}}\) and \(\refer{p} = \refer{\rho}R\refer{T}\) be
the reference length, density, temperature, velocity, and pressure, respectively.
The specific gas constant \(R = k_B/m\), where \(k_B\) is the Boltzmann constant, and \(m\) is the molar mass.
Then, \(f\refer{\rho}/\refer{c}^3\) is the one-particle velocity distribution function (VDF)
defined in seven-dimensional space \((tL/\refer{c}, \bx L, \bxi \refer{c})\) and
the macroscopic variables take the following form:
\(\rho\refer{\rho}\) is the density, \(\bv\refer{c}\) is the velocity, \(T\refer{T}\) is the temperature,
\(p_{\alpha\beta}\refer{p}\) is the stress tensor, \(\bq\refer{p}\refer{c}\) is the heat flux.
In the dimensionless form, they are calculated as polynomial moments of the VDF:
\begin{equation}\label{eq:macro}
    \begin{gathered}
    \rho = \int f \dxi, \quad
    \bv = \frac1{\rho} \int \bxi f \dxi, \quad
    T = \frac{1}{3\rho}\int|\bxi-\bv|^2f \dxi = \frac{p_{\alpha\alpha}}{3\rho}, \\
    p_{\alpha\beta} = \int(\xi_\alpha - v_\alpha)(\xi_\beta - v_\beta) f \dxi, \quad
    \bq = \frac12\int(\bxi-\bv)|\bxi-\bv|^2 f \dxi.
    \end{gathered}
\end{equation}
Integration with respect to \(\bxi\) is, hereafter, carried out over \(\mathbb{R}^3\).

%%% Boltzmann equation
The VDF is governed by the Boltzmann equation
\begin{equation}\label{eq:Boltzmann}
    \pder[f]{t} + \bxi\bdot\pder[f]{\bx} = \frac{1}{\tau} J(f),
\end{equation}
where \(J(f)\) is the collisional operator with a local Maxwellian as the equilibrium function
\begin{equation}\label{eq:equilibrium}
    \equil{f}\br{\bxi; \rho, \bv, T} = \frac{\rho}{(2\pi T)^{3/2}}\exp\br{-\frac{|\bxi-\bv|^2}{2T}}.
\end{equation}
The characteristic relaxation time of collisions to equilibrium \(\tau\)
can be expressed in terms of the reference gas viscosity \(\refer\mu\)~\cite{Cercignani2000, Sone2007},
\begin{equation}\label{eq:tau}
    \tau = \frac{\refer\mu\refer{c}}{\refer{p}L},
\end{equation}
and is related to the modified Knudsen number \(k=\ell_0\sqrt\pi/2L\), where \(\ell_0\) is the mean free path,
as \(\tau = k / \sqrt2\).

%%% BGK equation
In the present paper, we restrict ourselves to the simplest relaxation model~\cite{Krook1954, Welander1954}
\begin{equation}\label{eq:bgk_integral}
    J(f)(\bxi) = \equil{f}(\bxi) - f(\bxi),
\end{equation}
often referred as the Bhatnagar--Gross--Krook (BGK) model of the Boltzmann collisional operator.
The nonlinearity in~\eqref{eq:bgk_integral} is more severe in comparison to the full Boltzmann equation
since \(\equil{f}\) depends on \(f\) via its moments,
but the BGK model is much simpler from the numerical point of view.

%%% Boundary conditions
The gas--surface interaction is modeled via the diffuse-reflection boundary condition:
\begin{equation}\label{eq:diffuse}
    f(t,\bx_B,\bxi) = \equil{f} \br{ \bxi;
        -\sqrt\frac{2\pi}{T_B} \int_{\bxi'\bdot\bn<0} \br{ \bxi'\bdot\bn } f(t,\bx_B,\bxi')\dxi', \bv_B, T_B
    } \quad (\bxi\bdot\bn>0),
\end{equation}
where \(\bn\) is the unit vector normal to the boundary, directed into gas,
\(\bx_B\), \(T_B\), and \(\bv_B\) are the boundary coordinates, temperature, and velocity, respectively.
It is also assumed that \(\bv_B\bdot\bn = 0\).

\section{Discrete-velocity approximation}\label{sec:dv}

%%% Discretization of the velocity space
Within the DV framework, the admissible particle velocities
are restricted to the discrete set \(\Set{\bxi_j}{j=1,\dots,N}\).
Under this assumption, an arbitrary moment \(\phi(\bxi)\) of \(f\), including~\eqref{eq:macro},
is approximated as
\begin{equation}\label{eq:cubature}
    \int \phi f\dxi = \sum_{j=1}^{N} \phi(\bxi_j) f_j,
\end{equation}
where \(f_j/w_j\) is an approximation of \(f(\bxi_j)\), \(w_j\) is a quadrature weight.
The evolution of \(f_j\) is governed by the system of partial differential equations
\begin{equation}\label{eq:dvm}
    \pder[f_j]{t} + \bxi_j\bdot\pder[f_j]{\bx} = \frac{1}{\tau} J_N(f_j)
\end{equation}
which is called the DV model of~\eqref{eq:Boltzmann}~\cite{Cabannes1980}.
The \emph{DV method} of solving~\eqref{eq:Boltzmann} is based on a DV model,
which is consistent with~\eqref{eq:Boltzmann} when \(N\) goes to infinity~\cite{Aristov2001}.

%%% Conservative and entropic properties
It is important for a DV model~\eqref{eq:dvm} to preserve conservation and entropy properties
of the continuous kinetic equation~\eqref{eq:Boltzmann}.
For the BGK model
\begin{equation}\label{eq:dvm-bgk}
    J_N(f_j) = \equil{f}_j - f_j,
\end{equation}
it can be accomplished when the discrete local equilibrium \(\equil{f}_j\) has the Maxwellian form
\begin{equation}\label{eq:equilibrium-bgk}
    \equil{f}_{\DV,j}(\hat\bm) = \frac{\hat\rho w_{\DV,j}}{(2\pi \hat{T})^{3/2}}
        \exp\br{-\frac{|\bxi_{\DV,j}-\hat\bv|^2}{2\hat{T}}}, \quad
    \hat\bm = \transpose{\br{\hat\rho, \hat\bv, \hat{T}}} \in \mathbb{R}^5,
\end{equation}
where vector \(\hat\bm\) is the solution of
\begin{equation}\label{eq:m_solution}
    \sum_j \bpsi_{\DV,j} \br{ \equil{f}_{\DV,j}(\hat\bm) - f_j } = 0, \quad
    \bpsi_j = \transpose{\br{1, \bxi_j,| \bxi_j|^2}} \in \mathbb{R}^5.
\end{equation}
For a uniform lattice in the velocity space,
it is proved that there exists a unique discrete equilibrium~\eqref{eq:equilibrium-bgk},
which is the maximization of the discrete entropy functional
and guarantees that mass, momentum, and kinetic energy are conserved~\cite{Mieussens2000}.
For computations, a finite number of discrete velocities is used: \(N = N_\DV < \infty\).
Note also that moments of~\eqref{eq:equilibrium-bgk} are not associated with \(\hat\bm\):
\begin{equation}\label{eq:equilibrium-bgk-moments}
    \sum_j \bpsi_{\DV,j} \equil{f}_{\DV,j}(\hat\bm) = \rho\transpose{\br{1, \bv, |\bv|^2 + 3T}}, \quad
    \bm = \transpose{\br{\rho, \bv, T}} \in \mathbb{R}^5.
\end{equation}
Vector \(\bm\) is not equal to \(\hat\bm\), but is quite close to it in practice,
when the velocity grid is well suited to the problem.

%%% Anisotropic velocity grids
When the velocity grid is anisotropic, the diagonal terms of the pressure tensor \(p_{\alpha\beta}\)
calculated from \(\equil{f}_{\DV,j}\) are not in general equal to each other.
Hence, the discrete equilibrium~\eqref{eq:equilibrium-bgk} has at least a direction-dependent momentum flux
\(p_{\alpha\beta}n_\alpha \neq \rho T n_\beta\) across an arbitrary interface with unit normal \(\bn\).
Thus, 5-moment Maxwellian~\eqref{eq:equilibrium-bgk} is sufficient to construct the conservative DV method
even for nonuniform anisotropic grids in the velocity space, but not for hybrid numerical method
based on coupling of DV approximations with different degrees of \(p_{\alpha\beta}\) anisotropy.
This anisotropy can be eliminated if the discrete equilibrium is constructed in the form
\begin{equation}\label{eq:equilibrium-bgk-aniso}
    \equil{f}_{\DV,j}(\hat\bm) = \hat{\rho} w_{\DV,j} \prod_\alpha\br{ 2\pi \hat{T}_\alpha }^{-1/2}
        \exp\br{-\sum_\alpha\frac{\br{\xi_{\DV\alpha,j}-\hat{v}_\alpha}^2}{2\hat{T}_\alpha}}, \quad
    \hat\bm = \transpose{\br{\hat\rho, \hat\bv, \hat{\bT}}} \in \mathbb{R}^7,
\end{equation}
where \(\xi_{\DV\alpha,j} \equiv \bxi_{\DV,j}\), vector \(\hat\bm\) is the solution of
\begin{equation}\label{eq:m_solution-aniso}
    \sum_j \bpsi_{\DV,j} \br{ \equil{f}_{\DV,j}(\hat\bm) - f_j } = 0, \quad
    \bpsi_j = \transpose{\br{1, \bxi_j, \bxi_j^2}} \in \mathbb{R}^7.
\end{equation}
The 7-moment Maxwellian~\eqref{eq:equilibrium-bgk-aniso} can also be obtained
from generalized minimum entropy principle~\cite{Charrier1999}.

%%% LB method
The \emph{LB method} can be considered as a special discretization of the BGK model~\cite{Succi2001, Kruger2017}.
We assume that the considered flow is isothermal and slow, i.e., the Mach number is close to zero.
Then we can expand the local Maxwell state into the Taylor series on the bulk velocity $\bv$
and keep only the terms of some finite order (at least second).
Moreover, we assume that the particle can travel with the velocities $\bxi_{\LB,j}, j = 1 \ldots N_\LB$.
The values of absolute Maxwellian are changed by the lattice weights $w_{\LB,j}$ in such a way that the first moments
of the local equilibrium state are the same as for the Maxwell distribution.
For the LB models, the local equilibrium takes a polynomial form on the bulk velocity,
and the requirement of the conservation of mass, momentum, and energy yields the algebraic equations
for the lattice weights and velocities, which can be solved explicitly.
Therefore the conservation properties for LB method are achieved with much less efforts than for the conventional DV method.

%%% The equilibrium state
The third-order expansion in $\bv$ yields the following local equilibrium LB state:
\begin{equation}\label{eq:lbgk}
    \equil{f}_{\LB,j}(\bm) = \rho w_{LB,j}\left( 1
        + \frac{\bxi_{\LB,j}\bdot\bv}{c_s^2}
        + \frac{(\bxi_{\LB,j}\bdot\bv)^2-c_s^2v^2}{2c_s^4} + \frac{(\bxi_{\LB,j}\bdot\bv)^3-3c_s^2 v^2(\bxi_{\LB,j}\bdot\bv)}{6c_s^6}
    \right),
\end{equation}
where $c_s$ is the constant sound velocity defined by $\sum_jw_j\bxi^2_{\LB,j} = 3c_s^2$.
In the present study, the value of $c_s$ is calibrated to unity for all LB models.
The standard notation D3Q\(p\) means \(N_\LB=p\) for the three-dimensional LB model.
% The sound velocity $c_s$ is linearly dependent on $\Delta x/\Delta t$,
% where $\Delta x, \Delta t$ are the distance between the lattice  nodes and lattice time step respectively
% (for instance, for D3Q19 model one has $c_s=\sqrt{1/3} \Delta x/\Delta t$),
% therefore, $c_s$ can be calibrated by the choice of $\Delta x, \Delta t$,
% in the present study the value of $c_s$ equals unity is adopted for all the considered LB models.
In the case of low-order lattices like D3Q19, the third-order terms are truncated in \eqref{eq:lbgk}.
Hereinafter, a quadrature rule, based on \(\bxi_{\LB,j}\) and \(w_{\LB,j}\),
together with the discrete operator in form~\eqref{eq:lbgk}, is referred to as the LB model.

%%% Advantages of the LB method over the DV method
When the VDF is a slightly disturbed equilibrium,
it can be efficiently approximated using quadratures with a small number \(N_\LB\).
The LB models are capable of reproducing low-order polynomial moments of the VDF accurately and,
therefore, describing a fluid-dynamic behavior of a gas, including that beyond the NS level.
Several approaches can be applied for the construction of LB models
like Gauss--Hermite~\cite{He1997, Shan1998, Shan2006, Shan2010}
and the entropic method~\cite{Karlin1999, Chikatamarla2006, Chikatamarla2009}.

%%%%%%%%%%%%%%%%%%%%%%%%%%%%%%%
%%% Rogozin: We do not use this work, since we do not solve classical LBGK
%%%%%%%%%%%%%%%%%%%%%%%%%%%%%%%
%We apply the kinetic boundary conditions (diffuse reflection)~\cite{Ansumali2002} for LB models in the present paper.

%%% Boundary conditions
The diffuse-reflection boundary condition~\eqref{eq:diffuse} is discretized as
\begin{equation}\label{eq:diffuse-dv}
    f_j(t,\bx_B) = -\frac{\sum_{\bxi_k\bdot\bn<0} \br{ \bxi_k\bdot\bn } f_k(t,\bx_B)}
        {\sum_{\bxi_k\bdot\bn>0} \br{ \bxi_k\bdot\bn } \equil{f}_k(\bm_B)}
        \equil{f}_j(\bm_B) \quad \br{ \bxi_j\bdot\bn>0 }, \quad \bm_B = \transpose{\br{1, \bv_B, T_B}},
\end{equation}
which preserves conservation of mass.

\section{The mapping method}\label{sec:mapping}

We will introduce the mapping method in the spatial overlapping zone of the BGK and LB models.
First of all, we assume that in this domain the VDF of the gas is close to the Maxwell state
with zero bulk velocity and unit temperature.
Therefore, VDF can be represented in the form of the truncated Grad expansion
up to the third-order terms on the velocity
\begin{equation}\label{eq:grad}
    f_H(\bx,\bxi) = \omega(\bxi)\left(
        a(\bx) + \sum_{\alpha}\Aa(\bx)\Ha +
        \frac{1}{2!}\sum_{\alpha\beta}\Aab(\bx)\Hab +
        \frac{1}{3!}\sum_{\alpha\beta \gamma}\Aabg(\bx)\Habg
    \right),
\end{equation}
where $\Ha, \Hab, \Habg$ are the Hermite polynomials of the first, second, and third order
(in the case of low-order LB models only the terms up to the second order are retained).
The polynomials are defined by
\begin{equation}\label{eq:hermite}
    H_\alpha(\bxi) = \frac{(-1)}{\omega(\bxi)}\pder{\xi_\alpha}\omega(\bxi), \quad
    \Hab(\bxi) = \frac{1}{\omega(\bxi)}\pderder{\xi_\alpha}{\xi_\beta}\omega(\bxi), \quad
    \Habg(\bxi) = \frac{(-1)}{\omega(\bxi)}\pderderder{\xi_\alpha}{\xi_\beta}{\xi_\gamma}\omega(\bxi),
\end{equation}
and
\begin{equation}\label{eq:omega}
    \omega(\bxi) = \frac{1}{(2\pi)^{3/2}}\exp\br{ -\frac{\bxi^2}{2} }.
\end{equation}
The coefficients $a, \Aa, \Aab, \Aabg$ depend on $\bx$ (the point in the overlapping domain).
We will use the function \eqref{eq:grad} for the transfer of the data between the LB and the BGK models.

In the overlapping spatial domain, where two methods (LB and DV) are applied,
one needs to introduce the procedure of the data exchange between LB and DV methods.
It consists of two parts: the mapping from DV to LB and the mapping from LB to DV.
These steps are discussed below.

%At the first step, we update the DV VDF $f_\DV(x,\bxi)$ for the discrete velocities $\bxi_n$
%such that $(\bxi_n,\mathbf{e})<0$, where $\mathbf{e}$ is the outer normal to the overlapping domain.
%We start from the spatial nodes at the wall and move towards the overlapping zone.
{\bf DV to LB.}
Similarly to~\cite{Staso2016long}, this step can be termed as \emph{projection},
since the DV VDF, which contains information about all the moments,
is mapped on LB VDF, which describes correctly only some low-order moments.

Firstly, in the overlapping spatial domain (physical domain), we map the DV VDF on the Grad VDF
by calculating the following coefficients:
$$
   \begin{gathered}
        a=\sum_{j=1}^{N_{\DV}} f_{\DV,j}, \quad
        \Aa=\sum_{j=1}^{N_{\DV}}f_{\DV,j}\Ha(\bxi_{\DV,j}), \quad
        \Aab=\sum_{j=1}^{N_{\DV}}f_{\DV,j}\Hab(\bxi_{\DV,j}), \quad
        \Aabg=\sum_{j=1}^{N_{\DV}}f_{\DV,j}\Habg(\bxi_{\DV,j}).
    \end{gathered}
$$
%where $\bxi_{DV,j}, m=1 \ldots N_{DV}$ are the velocities of the DV difference scheme.
Now the Grad VDF \eqref{eq:grad} is recovered in the overlapping spatial domain.

Next, we will map~\eqref{eq:grad} on the LB distribution function using the Gauss--Hermite quadrature method.
The idea of the method is based on the fact that the representation of the VDF in the Grad form
is equivalent to the LB method~\cite{He1997, Shan1998, Shan2006}.
%We consider the first moments $a, \Aa, \Aab, \Aabg$ in the integral form
%and then calculate them using Gauss--Hermite quadratures
%\begin{equation}
%    \{ a, \Aa, \Aab, \Aabg \}=\int f(\bxi)\{ 1, \Ha, %\Hab, \Habg \}(\bxi)d\bxi =
%    \sum_{j=1}^N w_j\frac{f_H(\bc_j)}{\omega(\bc_j)} \{ %1, \Ha(\bc_j), \Hab(\bc_j), \Habg(\bc_j) \},
%\end{equation}
%where $w_j, \bc_j$ are the weights and the nodes of the %Gauss--Hermite quadrature respectively.
%The nodes $\bc_j$ can be considered as the LB %velocities,
%while $ w_j\frac{f(\bc_j)}{\omega(\bv_j)}$ are the LB %VDF values
%and $w_j$ are the LB analog of the Maxwell distribution.
Then the formula
\begin{equation}\label{eq:grad_to_latt}
    f_{\LB,j} = w_j\frac{f_H(\bxi_{\LB,j})}{\omega(\bxi_{\LB,j})}
\end{equation}
gives the mapping of $f_H$ to $f_{\LB,j}$.
% for the corresponding velocities $\bxi_{LB,j}, j=1 \ldots N_{LB}$.

{\bf LB to DV.}
Following~\cite{Staso2016long}, this step can be termed as \emph{reconstruction},
since one needs to find DV VDF having only the LB VDF,
which has less information about the moments than DV VDF.

Using LB VDF we evaluate the moments $a, \Aa, \Aab, \Aabg$  using the formulas
$$
\begin{gathered}
    a=\sum_{j=1}^{N_{\LB}} f_{\LB,j}, \quad
    \Aa=\sum_{j=1}^{N_{\LB}} f_{\LB,j}\Ha(\bxi_{\LB,j}), \quad
    \Aab=\sum_{j=1}^{N_{\LB}} f_{\LB,j}\Hab(\bxi_{\LB,j}), \quad
    \Aabg=\sum_{j=1}^{N_{\LB}} f_{\LB,j}\Habg(\bxi_{\LB,j})
\end{gathered}
$$
and recover Grad distribution function \eqref{eq:grad}.
Finally, the DV VDF is obtained by appropriate discretization of the Grad VDF.

%%% Extension to the regularized models
The described mapping method can be generalized for the LB models,
which are not derived from the Gauss--Hermite quadratures.
We assume that after the regularization procedure~\cite{Latt2006, Chen2006}
and~\cite{Zhang2006, Mont2015, Mattila2017},
the non-equilibrium part of LB VDF will be projected into a velocity space
with a basis spanned by Hermite polynomials of the appropriate order.
Then the equivalence between the LB VDF and the expansion of the Grad type can be achieved;
therefore, the proposed mapping method can be applied.

\section{Numerical method}\label{sec:numerics}

\subsection{Time-integration method}\label{sec:numerics:splitting}

%%% Splitting scheme
For the present study, we start from the straightforward numerical algorithm
providing the second-order accuracy for both time and physical coordinates.
Equation~\eqref{eq:Boltzmann} is solved by the symmetric Strang's splitting scheme
\begin{equation}\label{eq:Strang}
    S^{\Delta{t}}_{A+B}(f_0) = S^{\Delta{t}/2}_A \br{S^{\Delta{t}}_B \br{S^{\Delta{t}/2}_A(f_0)} } + \OO{\Delta{t}^3},
\end{equation}
where \(A(f) = -\bxi\bdot\nabla{f}\), \(B(f) = J(f)/\tau\), \(\Delta{t}\) is the time step.
\(S^t_P(f_0)\) denotes the solution of the Cauchy problem
\begin{equation}\label{eq:Cauchy}
    \pder[f]{t} = P(f), \quad f|_{t=0} = f_0.
\end{equation}
Scheme~\eqref{eq:Strang} preserves the second-order accuracy if \(S^t_A\) and \(S^t_B\)
are approximated with the same convergence properties individually~\cite{Bobylev2001}.

%%% Solutions of individual problems
An essential advantage of the splitting procedure is that the spatially homogeneous Boltzmann equation
has an exact solution for the BGK collision model:
\begin{equation}\label{eq:bgk_exact}
    S^t_B(f_0) = \equil{f} + \br{f_0-\equil{f}}\exp\br{-\frac\rho\tau t};
\end{equation}
however, the generalization to the original Boltzmann equation is straightforward.
For the collisionless Boltzmann equation, the desired second-order accuracy of \(S^t_A(f_0)\)
is archived by means of the finite-volume scheme described below.

%%% Steady-state solution
To find a steady-state solution of the boundary-value problem,
the time-marching process is started from some initial approximation
and continues until the convergence criterion is met.
For the DV method, the discrete equilibrium \(\equil{f}_{\DV,j}\) is obtained as a solution
of the nonlinear equations~\eqref{eq:m_solution} or \eqref{eq:m_solution-aniso} at each time step.
For this purpose, some root-finding algorithm should be harnessed.

\subsection{Finite-volume formulation}\label{sec:numerics:fv}

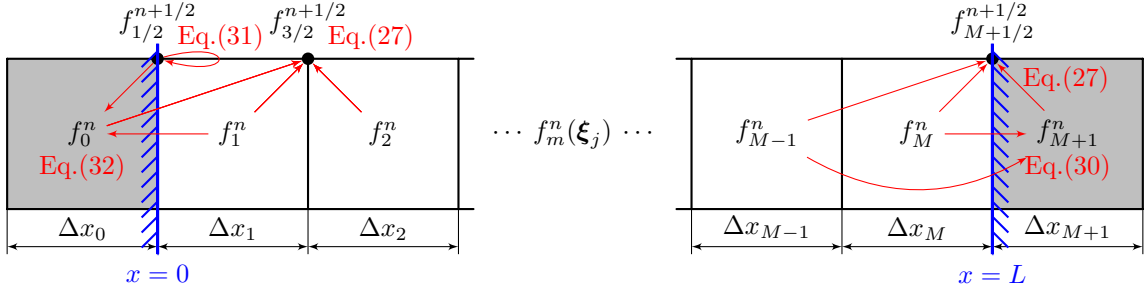
\begin{figure}
    \centering
    \begin{tikzpicture}
        \fill[ghost] (-2,2) -- (0,2) -- (0,0) -- (-2,0) -- cycle;
        \draw[thick, line cap=round, step=2] (-2,0) grid (4.2,2);
        \foreach \i [evaluate={\x=\i-1}] in {1,3}
            \node[bolddot, label=above:$f^{n+1/2}_{\i/2}$] (F\i) at (\x,2) {};
        \foreach \i [evaluate={\x=2*\i-1}, evaluate={\a=2*(\i-1)}, evaluate={\b=2*\i}] in {0,1,2} {
            \node (f\i) at (\x,1) {$f^n_{\i}$};
            \draw[<->] (\a,-.5) -- (\b,-.5) node[above=-1pt, pos=0.5] {$\Delta{x}_{\i}$};
        };
        \foreach \x in {-2,0,2,4} \draw (\x,0) -- (\x,-.6);
        \draw[boundary=-45](0,2.2) -- (0,-.6) node[below] {$x=0$};
        \foreach \i [] in {0,1,2}
        \foreach \n in {f0,f1,f2} \draw[->, eqref] (\n) -- (F3);
        \foreach \n in {F1,f1} \draw[->, eqref] (\n) -- (f0);
        \draw[eqref,->] (F1) to [loop right, scale=2] (F1);
        \node at (f0) [eqref, below=4pt] {Eq.\eqref{eq:first_ghost}};
        \node at (F1) [eqref, above=8pt, right=4pt] {Eq.\eqref{eq:first_flux}};
        \node at (F3) [eqref, above=8pt, right=4pt] {Eq.\eqref{eq:internal_fluxes}};
        \useasboundingbox (-2.2,-1.1) rectangle (4.2,2.8);
    \end{tikzpicture}
    \begin{tikzpicture}
        \node {$\cdots\:f^n_m(\bxi_j)\:\cdots$};
        \useasboundingbox (-0.1,-2.1) rectangle (0.1,1.8);
    \end{tikzpicture}
    \begin{tikzpicture}
        \fill[ghost] (0,2) -- (2,2) -- (2,0) -- (0,0) -- cycle;
        \draw[thick, line cap=round, step=2] (-4.2,0) grid (2,2);
        \node[bolddot, label=above:$f^{n+1/2}_{M+1/2}$] (F) at (0,2) {};
        \foreach \idx/\i [evaluate={\x=2*\i-1}, evaluate={\a=2*(\i-1)}, evaluate={\b=2*\i}] in {M-1/-1,M/0,M+1/1} {
            \node (f\i) at (\x,1) {$f^n_{\idx}$};
            \draw[<->] (\a,-.5) -- (\b,-.5) node[above=-1pt, pos=0.5] {$\Delta{x}_{\idx}$};
        };
        \foreach \x in {-4,-2,0,2} \draw (\x,0) -- (\x,-.6);
        \draw[boundary=45](0,2.2) -- (0,-.6) node[below] {$x=L$};
        \foreach \n in {f-1,f0,f1} \draw[->, eqref] (\n) -- (F);
        \draw[->, eqref] (f-1) to[bend right] (f1);
        \draw[->, eqref] (f0) -- (f1);
        \node at (f1) [eqref, below=4pt] {Eq.\eqref{eq:last_ghost}};
        \node at (F) [eqref, below=8pt, right=8pt] {Eq.\eqref{eq:internal_fluxes}};
        \useasboundingbox (-4.2,-1.1) rectangle (2.2,2.8);
    \end{tikzpicture}
    \caption{
        The computational domain \(\Omega=[0,L]\) is extended by two ghost cells (shown in gray).
        The blue lines highlight the boundaries of \(\Omega\).
        The red arrows represent the dependency relations
        among the set of nodes in the extended \(\Omega\) for \(\xi_j>0\).
        The values in nodes pointed out by these arrows are calculated from the provided references.
    }
    \label{fig:fv_scheme}
\end{figure}

%%% General form
For the sake of simplicity, we consider a one-dimensional physical space
and introduce \(\xi = \xi_1\), \(x = x_1\).
Then, the collisionless Boltzmann equation with discrete velocities
\begin{equation}\label{eq:transport}
    \pder[f(\bxi_j)]{t} + \xi_j\pder[f(\bxi_j)]{x} = 0
\end{equation}
is approximated by the finite-volume (FV) method:
\begin{equation}\label{eq:finite_volume}
    f^{n+1}_m = f^n_m - \gamma_m\br{f^{n+1/2}_{m+1/2} - f^{n+1/2}_{m-1/2}}, \quad
    \gamma_m(\bxi_j) = \frac{\xi_j\Delta{t}}{\Delta{x_m}}, \quad
    m = 1,\dots,M, \quad n\in\mathbb{N},
\end{equation}
where \(\Delta{x_m}\) is the width of \(m\) cell in the physical space (\(\bx\in\mathbb{R}\)),
\(f^n_m(\bxi_j)\) denotes the fully discretized VDF defined in the cell centers:
\begin{equation}\label{eq:vdf_fv}
    f^n_m(\bxi_j) = f\br{
        t = t_n \equiv n\Delta{t}, \;
        x = x_m \equiv \frac{\Delta{x_m}}2 + \sum_{l=1}^{m-1}\Delta{x_l}, \;
        \bxi = \bxi_j
    }
\end{equation}
and \(f^{n+1/2}_{m+1/2}(\bxi_j)\) (\(m = 0,\dots,M\)) are the reconstructed edge values
defined at \(x_{m+1/2} = \sum_{l=0}^m\Delta{x_l}\).
The computational domain \(\Omega = [0,L]\), where \(L = x_{M+1/2}\), is extended by two ghost cells,
which are corresponded to \(m=0\) and \(m=M+1\) (Fig.~\ref{fig:fv_scheme}).
Hereinafter, we describe only positive velocities \(\xi_j>0\).
Fig.~\ref{fig:fv_scheme} shows a mesh with two ghost cells at each boundary.
For \(\xi_j<0\), all expressions are analogous
and can be obtained by the replacement \(\xi_j\to-\xi_j\) and \(x\to-x\).

%%% Internal faces
The internal edge values can be written in the form
\begin{equation}\label{eq:internal_fluxes}
    f^{n+1/2}_{m+1/2} = f^n_m + \frac{1-\gamma_m}2\overline{Df^n_m}\Delta{x_m},
    \quad m = 1,\dots,M,
\end{equation}
where \(\overline{Df^n_m}\) is the limited approximation of \(\Pder[f]{x}(t_n, x_m, \bxi_j)\).
A monotonic-preserve scheme should be employed because sharp variations (in physical space)
of solution can occur even for nearly incompressible flow, especially for large \(|\bxi_j|\).
In the present paper, the third-order total variation diminishing (TVD) scheme is used
(see~\ref{sec:limiter} for details):
\begin{equation}\label{eq:limiter}
    \overline{Df^n_m} = \begin{cases}
        \min\br{
             \frac2{\gamma_m}\frac{|\Delta_-|}{h_-},
             \frac{1+\gamma_m}3\frac{|\Delta_-|}{h_-} + \frac{2-\gamma_m}3\frac{|\Delta_+|}{h_+},
             \frac2{1-\gamma_m}\frac{|\Delta_+|}{h_+}
        }\sgn\br{\Delta_-}, &\quad \Delta_+\Delta_- > 0, \\
        0, &\quad \Delta_+\Delta_- \leq 0,
    \end{cases}
\end{equation}
where
\begin{equation}\label{eq:differences}
    \Delta_\pm = \pm\br{f_{m\pm1}^n - f_m^n}, \quad h_\pm = \frac{\Delta{x_{m\pm1}} + \Delta{x_m}}2.
\end{equation}

%%% Boundary conditions
The outflow boundary condition are expressed in terms of edge values \(f^{n+1/2}_{M+1/2}(\bxi_j)\),
which are obtained from~\eqref{eq:internal_fluxes}
and the linear extrapolation for the ghost cell \(m=M+1\):
\begin{equation}\label{eq:last_ghost}
    f^n_{M+1} = 2f^n_M - f^n_{M-1}, \quad \Delta{x}_{M+1} = \Delta{x}_{M-1},
\end{equation}
which preserve the second-order accuracy~\cite{LeVeque2002}.
The inflow boundary condition is included into the scheme
by means of \(f^{n+1/2}_{1/2}\) and \(f^{n+1/2}_{3/2}\).
In particular, the DV diffuse-reflection boundary condition~\eqref{eq:diffuse-dv} yields
\begin{equation}\label{eq:first_flux}
    f^{n+1/2}_{1/2}(\bxi_j) = -\frac{\sum_{\xi_k>0} \xi_k f^{n+1/2}_{1/2}(\bxi_k)}
        {\sum_{\xi_k<0} \xi_k \equil{f}_k(\bm_B)}
        \equil{f}_j(\bm_B).
\end{equation}
For the ghost cell \(m=0\), used for calculation \(f^{n+1/2}_{3/2}\) via~\eqref{eq:internal_fluxes},
the linear extrapolation of the form
\begin{equation}\label{eq:first_ghost}
    f^n_0 = \frac2{1+\gamma_1}f^{n+1/2}_{1/2} - \frac{1-\gamma_1}{1+\gamma_1}f^n_1, \quad
    \Delta{x}_0 = \Delta{x}_1
\end{equation}
also preserves the second-order accuracy, since it is a linear extrapolation of two values at \(t=t_n\):
\(f^n_1\) and \(f^{n+1/2}_{1/2} \equiv f(t_n+\Delta{t}/2, 0, \bxi_j) = f(t_n, -\gamma_1\Delta{x_1}/2, \bxi_j)\).
It is worth emphasizing that the presented numerical scheme for the boundary-value problem
possesses the second-order accuracy along with conservation of mass
(in contrast to the FV scheme used in~\cite{Baranger2019}).

%%% Stability
The maximum value of the time step \(\Delta{t}\) for the described explicit scheme
is limited by the Courant--Friedrichs--Lewy (CFL) condition:
\(\gamma_m(\bxi_j)\leq1\) for all \(m\) and \(j\), which is equivalent to
\begin{equation}\label{eq:CFL}
    \Delta{t} \leq \min_{m,j} \frac{\Delta{x_m}}{\xi_j}.
\end{equation}

%%% Why half-integer lattice?
The boundary conditions also dictate a way of discretization in the velocity space.
Concerning the origin of the velocity coordinates, only two types of lattices are symmetric~\cite{Inamuro1990}:
integer \(\br{\bxi_j/c_s} \in \mathbb{Z}^3\) and half-integer \(\br{\bxi_j/c_s + \be/2} \in \mathbb{Z}^3\),
where \(\be\) is the corresponding orthonormal basis.
For the considered boundary condition at \(x=0\), there is a zero-measure set of velocities
\(\Set{\bxi\in\mathbb{R}^3}{\xi_1=0}\), called tangential.
These velocities are immune to the boundary conditions.
The integer lattice contains a substantial subset of tangential velocities.
Therefore, to avoid an additional discretization error, the half-integer lattice is employed.

%%% Diffuse reflection for LB
In the same manner, LB quadratures without tangential velocities are preferable to the classical ones.
The LB models can be supplemented by special velocity groups to more accurately approximate
the diffuse-reflection boundary condition by minimizing errors at the half-space moments~\cite{Feuchter2016}.
Moreover, the Gauss--Laguerre quadratures are able to reproduce the Maxwell half-range moments
exactly~\cite{Ambrus2014, Ambrus2016}.

\subsection{Coupling algorithm}\label{sec:numerics:coupling}

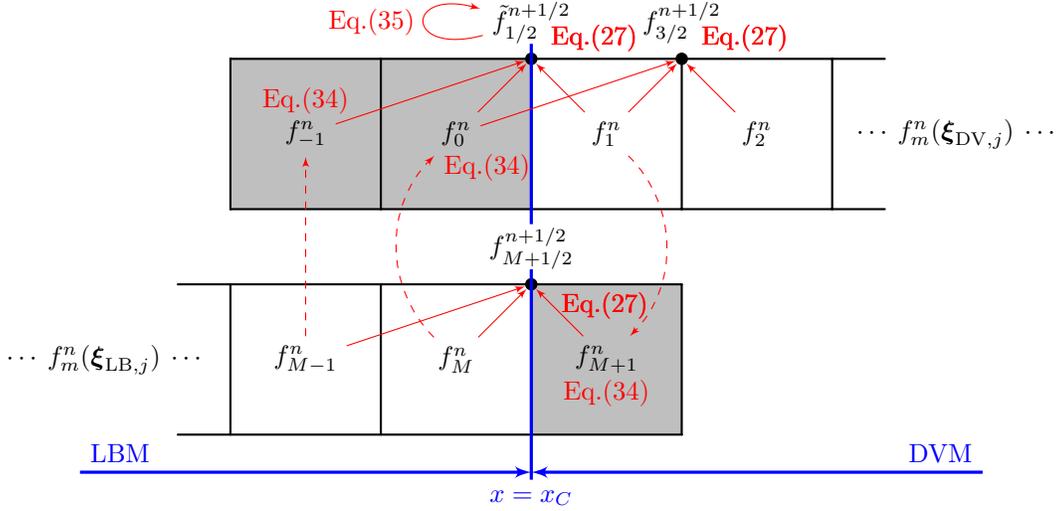
\begin{figure}
    \centering
    \begin{tikzpicture}
        \fill[ghost] (-4,2) -- (0,2) -- (0,0) -- (-4,0) -- cycle;
        \fill[ghost] (0,-1) -- (2,-1) -- (2,-3) -- (0,-3) -- cycle;
        \draw[thick, line cap=round, step=2] (-4,0) grid (4.7,2);
        \draw[thick, line cap=round, step=2, yshift=1cm] (-4.7,-4) grid (2,-2);
        \node[bolddot] (F_DV) at (0,2) {};
        \node[above] (F) at (F_DV.north) {$\tilde{f}^{n+1/2}_{1/2}$}
            edge [eqref, loop left] node {Eq.\eqref{eq:poly_correction}} ();
        \node[bolddot, label=above:$f^{n+1/2}_{3/2}$] (F_DV3) at (2,2) {};
        \node[bolddot, label=above:$f^{n+1/2}_{M+1/2}$] (F_LB) at (0,-1) {};
        \node[left](f_LB-2) at (-4.2,-2) {$\cdots\:f^n_m(\bxi_{\LB,j})\:\cdots$};
        \node[right](f_DV3) at (4.2,1) {$\cdots\:f^n_m(\bxi_{\DV,j})\:\cdots$};
        \foreach \idx\i [evaluate={\x=2*\i-1}] in {M-1/-1,M/0,M+1/1} {
            \node (f_DV\i) at (\x,1) {$f^n_{\i}$};
            \node (f_LB\i) at (\x,-2) {$f^n_{\idx}$};
        };
        \node (f_DV2) at (3,1) {$f^n_2$};
        \draw[interface](0,2.2) -- (0,-0.2);
        \draw[interface](0,-.8) -- (0,-3.6) node[below] {$x=x_C$};
        \draw[->, interface] (-6,-3.5) node[above right] {LBM} -- (0,-3.5);
        \draw[<-, interface] (0,-3.5) -- (6,-3.5) node[above left] {DVM};
        \foreach \m in {LB,DV} \foreach \n in {f_\m-1,f_\m0,f_\m1} {
            \draw[->, eqref] (\n) -- (F_\m);
        }
        \foreach \m in {LB} \foreach \n in {f_\m-1,f_\m0,f_\m1} {
            \node at (F_\m) [eqref, below=8pt, right=8pt] {Eq.\eqref{eq:internal_fluxes}};
        }
        \foreach \m in {DV} \foreach \n in {f_\m-1,f_\m0,f_\m1} {
            \node at (F_\m) [eqref, above=8pt, right=4pt] {Eq.\eqref{eq:internal_fluxes}};
        }
        \foreach \m in {DV} \foreach \n in {f_\m0,f_\m1,f_\m2} {
             \draw[->, eqref] (\n) -- (F_\m3);
            \node at (F_\m3) [eqref, above=8pt, right=4pt] {Eq.\eqref{eq:internal_fluxes}};
        }
        \draw[->, mapping] (f_LB-1) -- (f_DV-1);
        \foreach \a/\b in {LB0/DV0,DV1/LB1}
            \draw[->, mapping] (f_\a) to[bend left=45] (f_\b);
        \foreach \n/\side/\offset in {f_DV-1/above/0,f_DV0.west/below right/1mm, f_LB1/below/0}
            \node at ($(\n) - (\offset,0)$) [eqref, \side=4pt] {Eq.\eqref{eq:discrete_grad}};
        \useasboundingbox (-4.2,-1.1) rectangle (2.2,2.8);
    \end{tikzpicture}
    \caption{
        The computational domain \(\Omega\) is divided into LB and DV subdomains,
        which are extended by the ghost cells (shown in gray).
        The vertical blue line highlight the coupling interface \(x=x_C\).
        The red arrows represent the dependency relations
        among the set of nodes in the extended \(\Omega\) for \(\xi_j>0\).
        The values in nodes pointed out by these arrows are calculated from the provided references.
        The dashed red arrows correspond to the projection of the solution onto the truncated Hermite basis.
    }
    \label{fig:coupling_scheme}
\end{figure}

%%% Domain decomposition
The mapping approach presented in Sec.~\ref{sec:mapping} can be implemented within the FV framework.
Let us divide our physical domain into the LB and DV subdomains.
The coupling condition at the interface between these subdomains can be represented as a virtual boundary condition.
The proposed coupling strategy is based on the concept of ghost cells,
in which the VDF is reconstructed from the truncated Hermite expansion~\eqref{eq:grad}:
\begin{equation}\label{eq:discrete_grad}
    \hermite{f}_j = \equil{f}_j( \bm_0 ) \br{ a
        + \sum_{\alpha} \Aa \Ha(\bxi_j)
        + \frac12 \sum_{\alpha,\beta} \Aab \Hab(\bxi_j)
        + \frac16 \sum_{\alpha,\beta,\gamma} \Aabg \Habg(\bxi_j)
    },
\end{equation}
where \(\bm_0 = \transpose{\br{1, \boldsymbol{0}, 1}}\),
\(\equil{f}_{\LB,j}(\bm_0) = w_{\LB,j}\), which is seen from~\eqref{eq:lbgk},
and \(\equil{f}_{\DV,j}(\bm_0)\) is close to \(w_{\DV,j}\omega(\bxi_{\DV,j})\),
but not equal to it for the conservative DV method.

%%% 1D example
Let us return to the one-dimensional case described in Sec.~\ref{sec:numerics:fv}
and suppose that \(x=x_C\) is the coordinate of the coupling interface,
\(x<x_C\) and \(x>x_C\) are covered by the LB and DV models, respectively (Fig.~\ref{fig:coupling_scheme}).
As before, only the positive velocities \(\xi_j>0\) are described.
Negative velocities \(\xi_j<0\) are treated in the same manner due to
symmetric coupling based on the unified DV representation in the velocity space.
In order to use~\eqref{eq:internal_fluxes} for calculating \(f^{n+1/2}_{1/2}(\bxi_{\DV,j})\)
and \(f^{n+1/2}_{3/2}(\bxi_{\DV,j})\),
the VDF should be mapped from LB to DV model in the ghost cells \(m=-1\) and \(m=0\).
Conversely, \(f^n_{M+1}(\bxi_{\LB,j})\) is reconstructed from \(f^n_1(\bxi_{\DV,j})\)
and substituted in~\eqref{eq:internal_fluxes} to obtain \(f^{n+1/2}_{M+1/2}(\bxi_{\LB,j})\).

%%% Conservative scheme
The proposed Hermite-based mapping method preserves all the conservation properties
because all moments required for the equilibrium function are calculated precisely.
However, the FV scheme deals separately with velocities
directed in the opposite half-spaces with respect to the coupling interface.
For this reason, mass, momentum, and energy fluxes across the coupling interface
depend on quadrature rule and, therefore, are not exactly equal for the LB and DV models.
In the present study, the polynomial correction (like in~\cite{Aristov1980})
of the flux through the coupling interface is employed to recover the conservation properties.
Specifically, edge values \(f^{n+1/2}_{1/2}(\bxi_{\DV,j})\) are replaced by the corrected ones
\begin{equation}\label{eq:poly_correction}
    \tilde{f}^{n+1/2}_{1/2}(\bxi_{\DV,j}) = f^{n+1/2}_{1/2}(\bxi_{\DV,j}) \br{ 1 + \bm_C\bdot\bpsi_{\DV,j} }
        \quad (\xi_{\DV,j} > 0),
\end{equation}
where \(\bpsi_j\) is defined in~\eqref{eq:m_solution} and \(\bm_C\) is found from
\begin{equation}\label{eq:m_correction}
    \sum_{j=1}^{N_{\DV}} \xi_{\DV,j} \bpsi_{\DV,j} \tilde{f}^{n+1/2}_{\DV,j} =
        \sum_{j=1}^{N_{\LB}} \xi_{\LB,j} \bpsi_{\LB,j} f^{n+1/2}_{\LB,j}.
\end{equation}
For negative velocities (\(\xi_{\DV,j} < 0\)), the DV solution remains unchanged as well as the LB one.
In practice, each component of \(\bm_C\in\mathbb{R}^5\) is significantly less than unity;
therefore, the positivity is also preserved.

\subsection{Mesh refinement}\label{sec:numerics:refinement}

The diffuse-reflection boundary condition introduces several singularities into the VDF
both in the velocity space and in the physical one.
First, the discontinuity exists along the plane \(\bxi\bdot\bn=0\) and directly on the boundary surface (\(\bx=\bx_B\)).
For a convex domain, this discontinuity does not enter the gas region~\cite{Kim2011, Guo2017},
since characteristics do not enter into the gas region.
Nevertheless, sharp variations of the solution near \(\bxi\bdot\bn=0\) requires a strong mesh refinement
in order to achieve a high-accuracy approximation.
Moreover, it is necessary to resolve the logarithmic singularity of the form~\cite{Takata2016}
\begin{equation}\label{eq:xi_singularity}
    \pder[f]{\bxi}\bdot\bn = C_1\log\bxi\bdot\bn + \OO{1}, \quad (\bxi\bdot\bn<0, \quad \bx=\bx_B).
\end{equation}
Second, another logarithmic singularity arises in the physical space along \(\bn\)~\cite{Takata2014}:
\begin{equation}\label{eq:x_singularity}
    \pder[f]{\bx}\bdot\bn = \frac{C_2}{k}\log\frac{(\bx-\bx_B)\bdot\bn}{k} + \OO{1}.
\end{equation}
Here, \(C_1\) and \(C_2\) are some positive constants.
It is seen from~\eqref{eq:x_singularity} that the logarithmic singularity of the VDF
--- and all macroscopic variables as well --- takes place for all Knudsen numbers;
however, for small \(k\), it becomes highly localized in the Knudsen layer.
Therefore, the physical mesh should be refined exponentially as \(\bx\) goes to \(\bx_B\).

\section{Results and discussion}\label{sec:results}

\subsection{Couette-flow problem}

%%% Couette-flow problem & benchmark
The proposed numerical method is tested for the plane Couette-flow problem,
where a gas is embraced between the two parallel plates with non-zero relative velocity.
The hybrid approach assumes that a highly nonequilibrium gas in the Knudsen layer
is described using the BGK equation, while the LB model is employed for the internal zone.
All the presented results can be reproduced using python code~\cite{Rogozin2019hybrid}.

%%% Benchmark solutions
For the BGK model of the Boltzmann equation, the plane Couette-flow problem can be reduced
to the one-dimensional Fredholm integral equation with a weakly singular kernel
(see~\ref{sec:benchmark_solution} for details),
which has been solved accurately in~\cite{Luo2015} and, especially, in~\cite{Luo2016}.
Due to the lack of data on longitudinal heat flux in the mentioned works,
we have re-implemented~\cite{Rogozin2019couette}
the adaptive collocation method based on the generalized Gauss quadratures presented in~\cite{Luo2016}
for computing the benchmark solutions.

%%% Physical mesh
Let the plates be placed at \(y = \pm 1/2\) with constant temperature \(T = 1\)
and velocities (\(\pm\Delta v/2,0,0\)), where \(\Delta v=0.02\),
which is small enough to consider the Couette-flow problem as linear.
A completely diffuse reflection is assumed at the plates.
The average density is equal to unity: \(\int_{-1/2}^{1/2}\rho dy=1\).
The physical space \(0 < y < 1/2\) is divided in two subdomains: \(0 < y < y_I\) and \(y_I < y < 1/2\).
The first one consists of identical cells,
while the nonuniform mesh in the second one refines as a geometrical sequence near \(y = 1/2\).
The total number of cells depends on \(k\) and varies from 28 to 35.

%%% Velocity grid
The VDF in the velocity space varies from the discontinuous sum of two half-Maxwellians at the boundary
with complete diffuse-reflection condition to the near-equilibrium form in the vicinity of \(y=0\).
Such diversity can be efficiently approximated under the fixed DV set by employing
a significantly nonuniform velocity grid with local refinement near \(\xi_y=0\)~\cite{Ohwada1990, Wu2014, Rogozin2016}.
In the present paper, the nonuniform Cartesian lattice is cut off by the sphere of radius \(\xi^\cut=4\).
Along \(\xi_x\) and \(\xi_z\) axis, the nodes are distributed as the scaled roots of the Hermite polynomials:
\begin{equation}\label{eq:vgrid_hermite}
    \xiaj = x_{\Herm j} R, \quad w_{\alpha,j} = w_{\Herm j}\exp\br{x_{\Herm j}^2}R, \quad
    R = 2\xi^\cut \br{ \sum_{i=1}^{2N_{\alpha}} w_{\Herm i} \exp\br{x_{\Herm i}^2} }^{-1}, \quad
    j = 1,\dots,2N_{\alpha},
\end{equation}
where \(x_{\Herm j}\) and \(w_{\Herm j}\) are nodes and weights of the \((2N_{\alpha})\)-point Gauss--Hermite quadrature.
Along \(\xi_y>0\) semiaxis, the nodes are distributed as a polynomial (particularly, quadratic \(p=2\)) sequence:
\begin{equation}\label{eq:vgrid_polynomial}
    \xi_{y,j} = \sum_{i=0}^{j-1} \frac{w_{y,i} + w_{y,i+1}}2, \quad
    w_{y,j} = w_{y,1} + (j-1)^p \br{\xi^\cut - N_y w_{y,1}} \br{\sum_{i=0}^{N_y-1} i^p}^{-1}, \quad
    j = 1,\dots,N_y,
\end{equation}
where \(w_{y,0}=0\), \(w_{y,1}=0.1\) is the minimal width of the cell in the velocity space.
The quadrature weight in \(\mathbb{R}^3\) of discrete velocity \(\bxi_j\)
is equal to \(\prod_\alpha w_{\alpha,j}\).
For all the presented DVM solutions, \(N_{x/z}=8\) and \(N_y=16\) are chosen.
The discrete velocities with \(|\bxi_j| > \xi^\cut\) are excluded from the velocity grid;
therefore, the total number of discrete velocities \(N_\DV=5928\).

%%% Preliminary discussion on longitudinal heat flux
Here it is important to mention that the longitudinal heat flux $q_x$ exists in the linear Couette-flow problem,
or to be exact, in the Knudsen layer only and, therefore, decays exponentially with the distance from the boundary~\cite{Sone2007}.
This heat flux is caused by the diffuse-reflection boundary condition and has non-polar singularity on $k$ at the boundary.
Hence, it cannot be reproduced by the Navier--Stokes or any other high-order fluid-dynamic-type equations
obtained from the Hilbert or Chapman--Enskog expansions.
The same conclusion holds for the classical LB models.

\subsection{Numerical analysis}

\begin{figure}
    \centering
    \begin{subfigure}[b]{0.5\textwidth}
        \includegraphics[width=\textwidth]{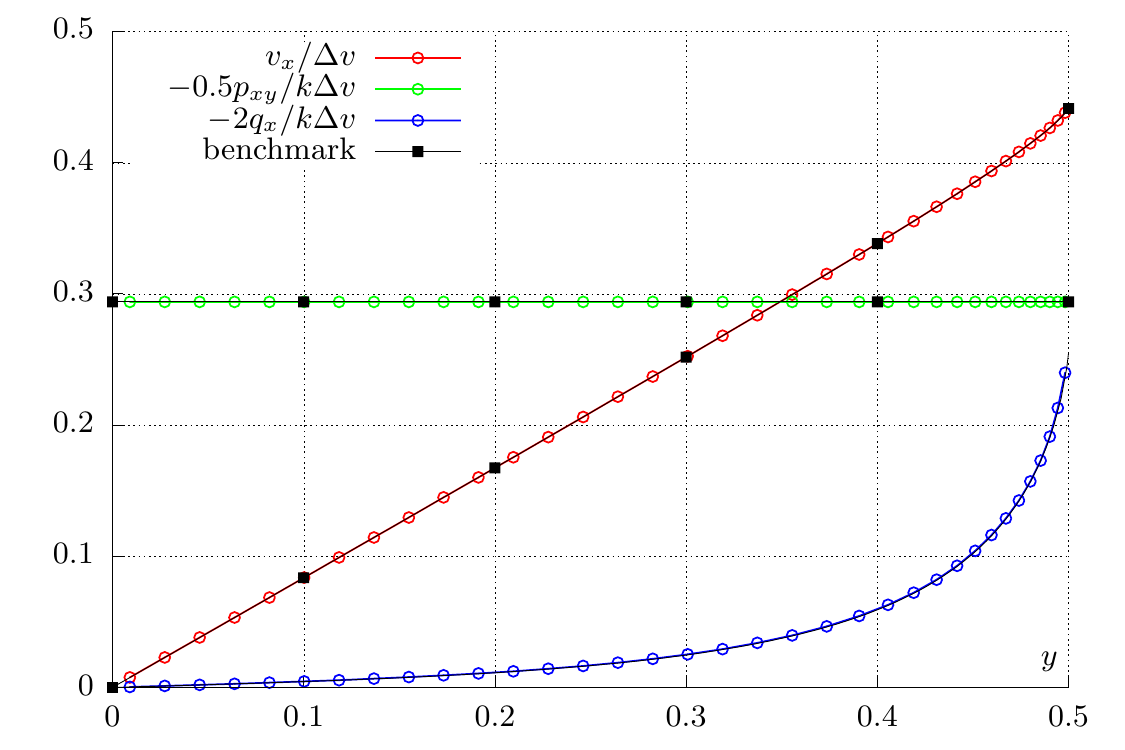}
        \caption{DV method}
        \label{fig:dvm}
    \end{subfigure}%
    \begin{subfigure}[b]{0.5\textwidth}
        \includegraphics[width=\textwidth]{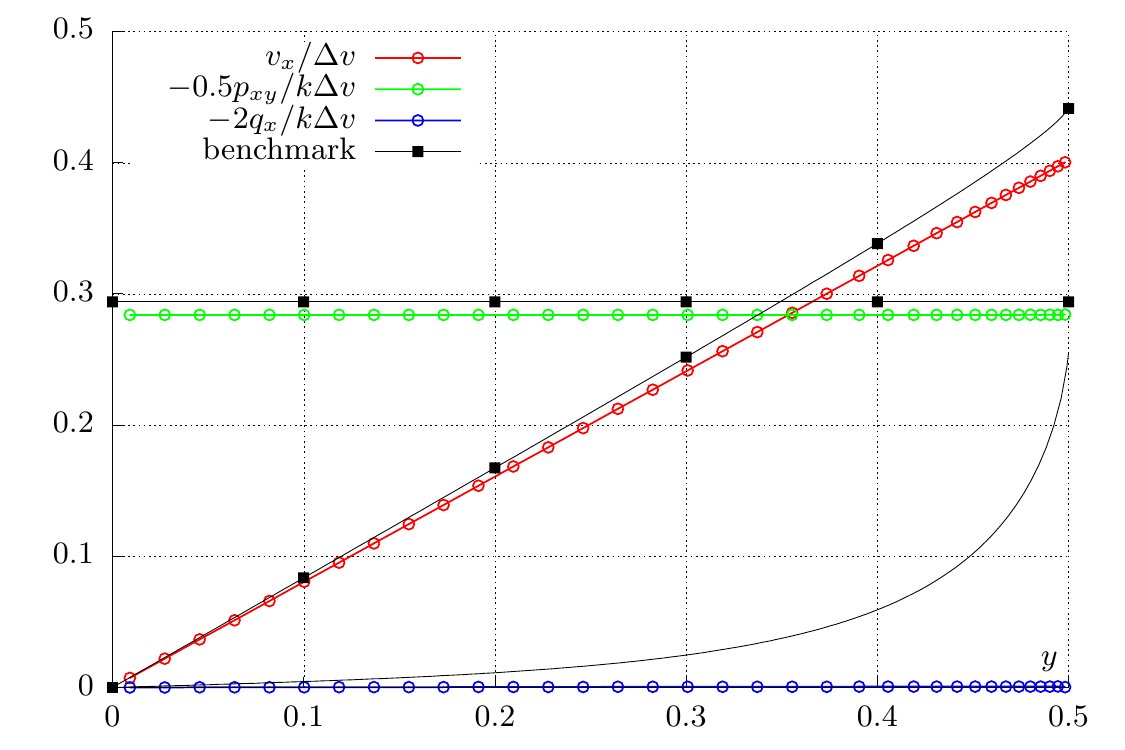}
        \caption{LB method: D3Q19}
        \label{fig:d3q19}
    \end{subfigure}\\
    \begin{subfigure}[b]{0.5\textwidth}
        \includegraphics[width=\textwidth]{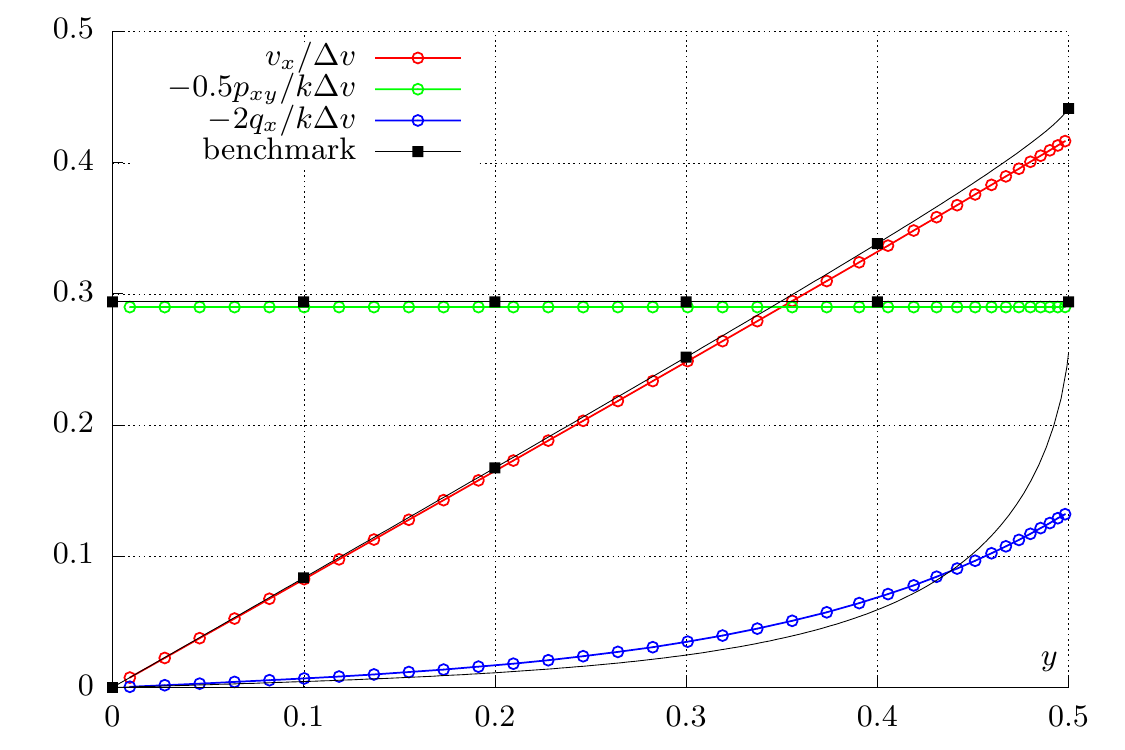}
        \caption{LB method: D3Q39}
        \label{fig:d3q39}
    \end{subfigure}%
    \begin{subfigure}[b]{0.5\textwidth}
        \includegraphics[width=\textwidth]{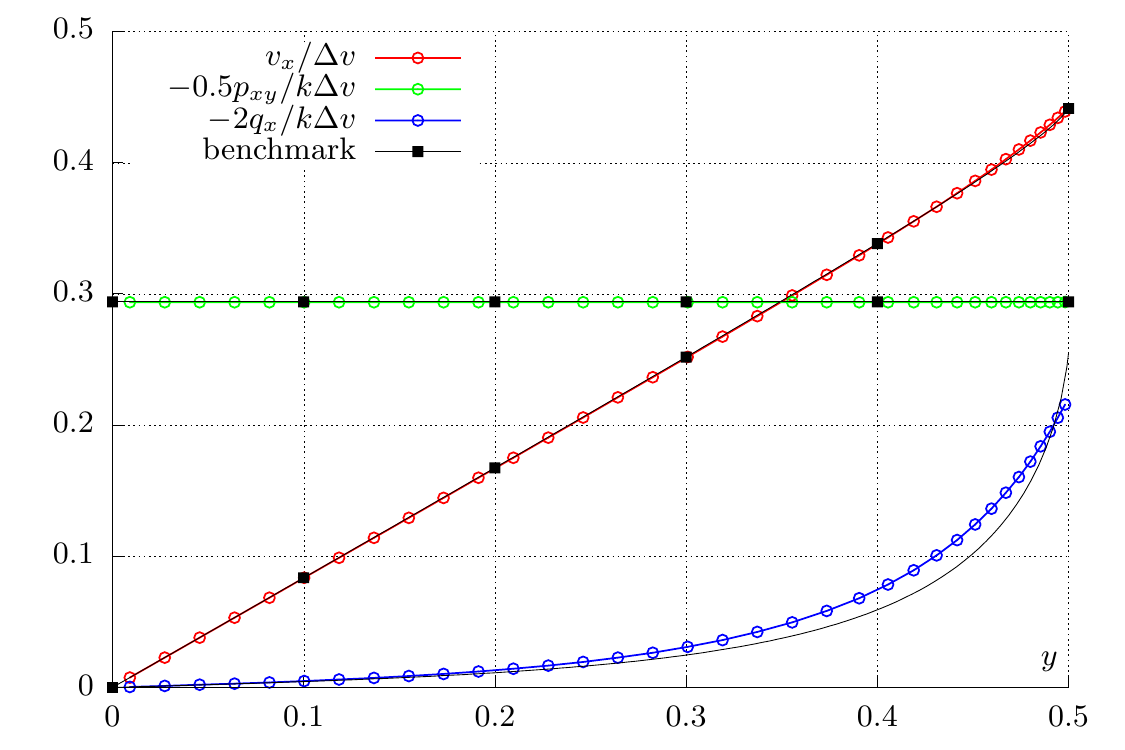}
        \caption{LB method: D3Q96}
        \label{fig:d3q96}
    \end{subfigure}
    \caption{
        Numerical solution of the Couette-flow problem for \(k=0.1\) obtained by pure DV or LB methods.
        The black lines are the high-accuracy solution for the BGK model.
        The black boxes correspond to the tabulated solutions~\cite{Luo2016}.
    }\label{fig:pure}
\end{figure}

\begin{figure}
    \centering
    \begin{subfigure}[b]{0.5\textwidth}
        \includegraphics[width=\textwidth]{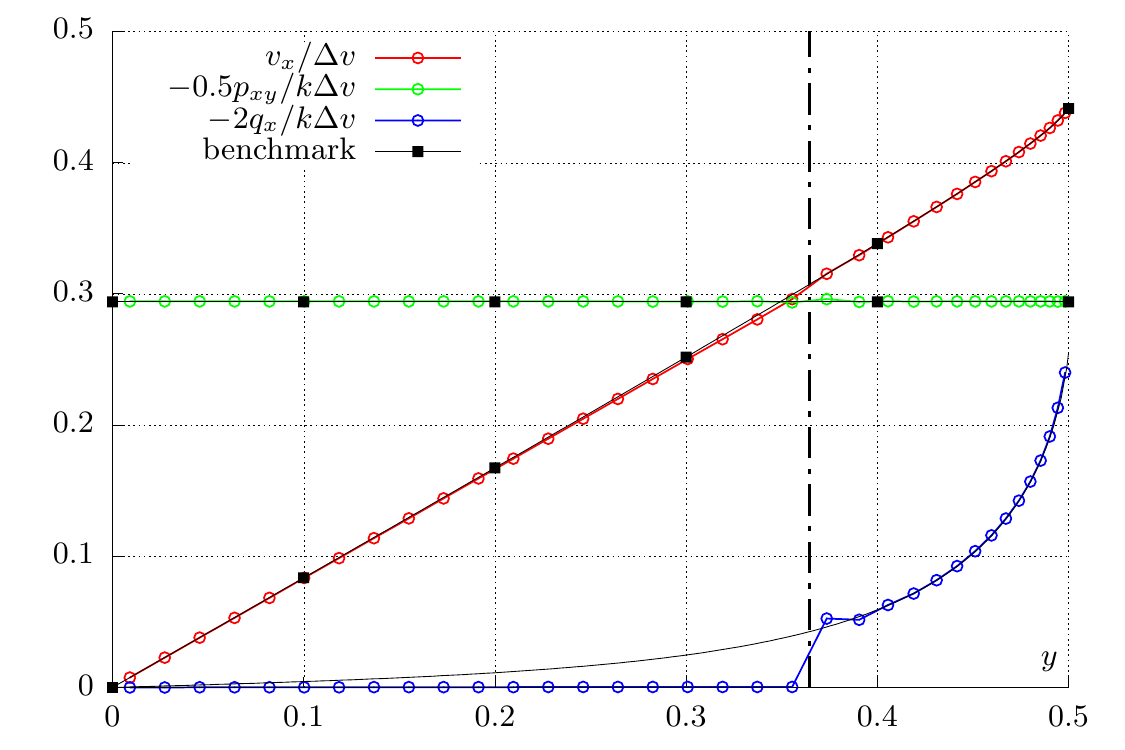}
        \caption{hybrid: DV and D3Q19}
        \label{fig:hyb:d3q19}
    \end{subfigure}%
    \begin{subfigure}[b]{0.5\textwidth}
        \includegraphics[width=\textwidth]{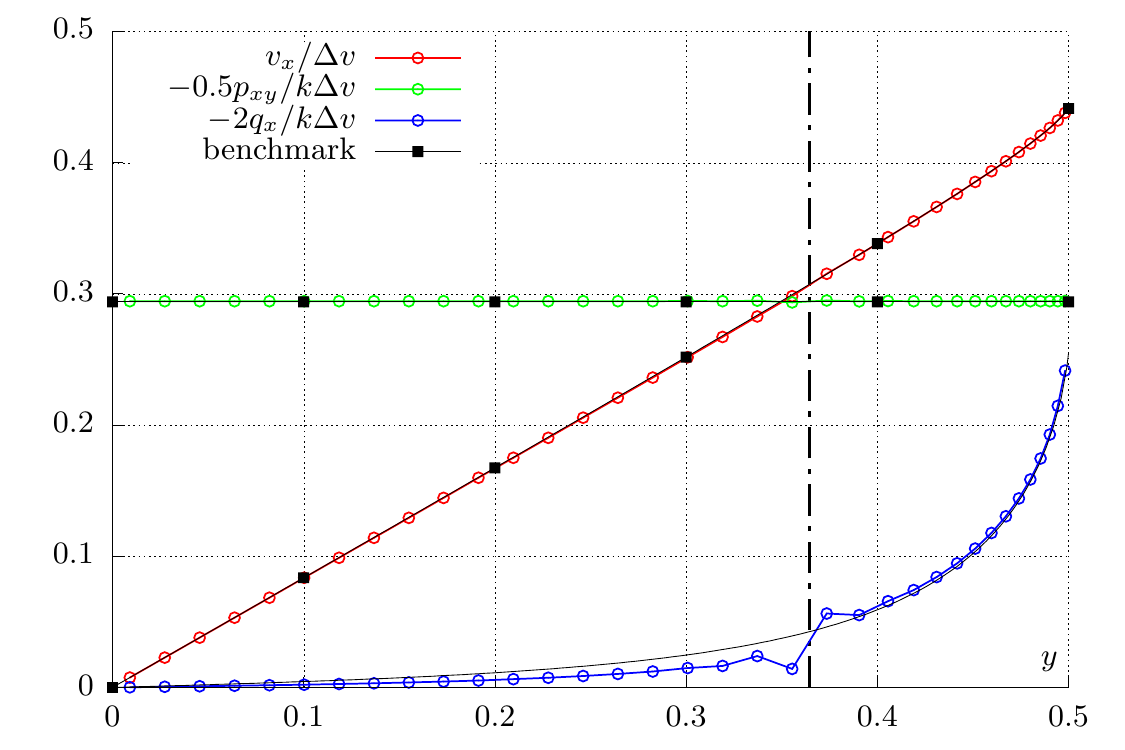}
        \caption{hybrid: DV and D3Q96}
        \label{fig:hyb:d3q96}
    \end{subfigure}
    \caption{
        Numerical solution of the Couette-flow problem for \(k=0.1\) obtained by the proposed hybrid method.
        There are \(1.2\) mean free paths between the boundary and coupling interface marked with the dash-dotted line.
        The black lines are the high-accuracy benchmark solution.
        The black boxes correspond to the tabulated values from~\cite{Luo2016}.
    }\label{fig:hybrid}
\end{figure}

\begin{figure}
    \centering
    \begin{subfigure}[b]{0.5\textwidth}
        \includegraphics[width=\textwidth]{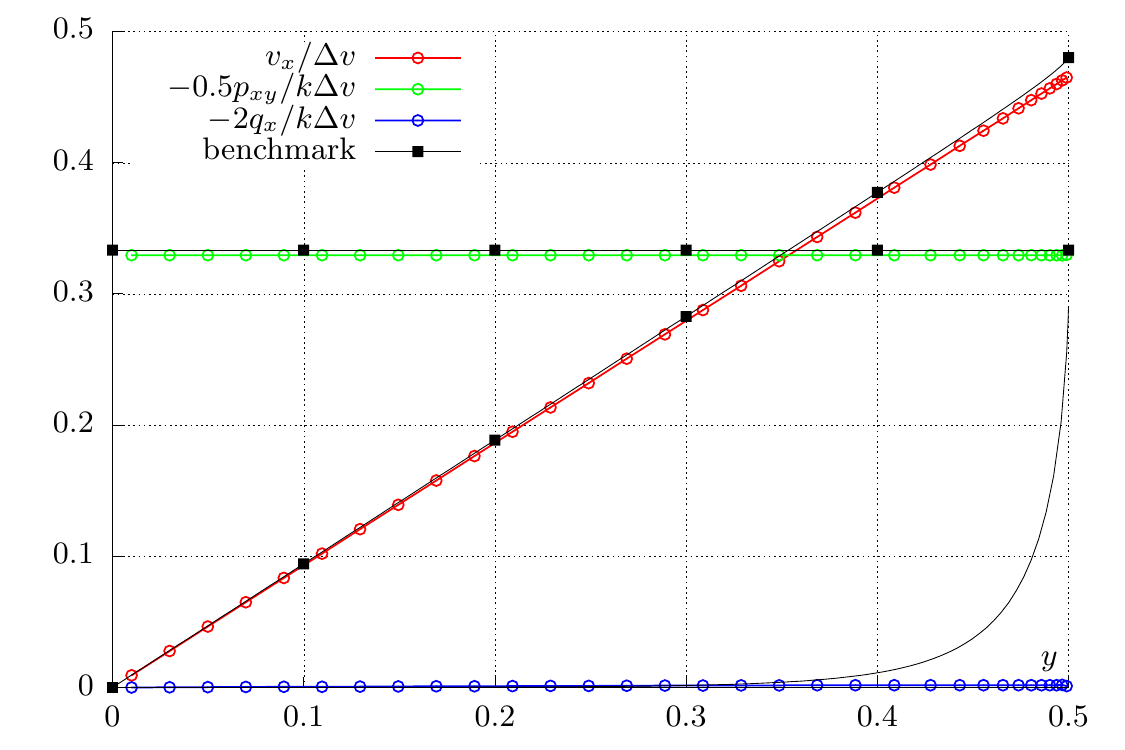}
        \caption{LB method: D3Q19}
        \label{fig:d3q19-k0_03}
    \end{subfigure}%
    \begin{subfigure}[b]{0.5\textwidth}
        \includegraphics[width=\textwidth]{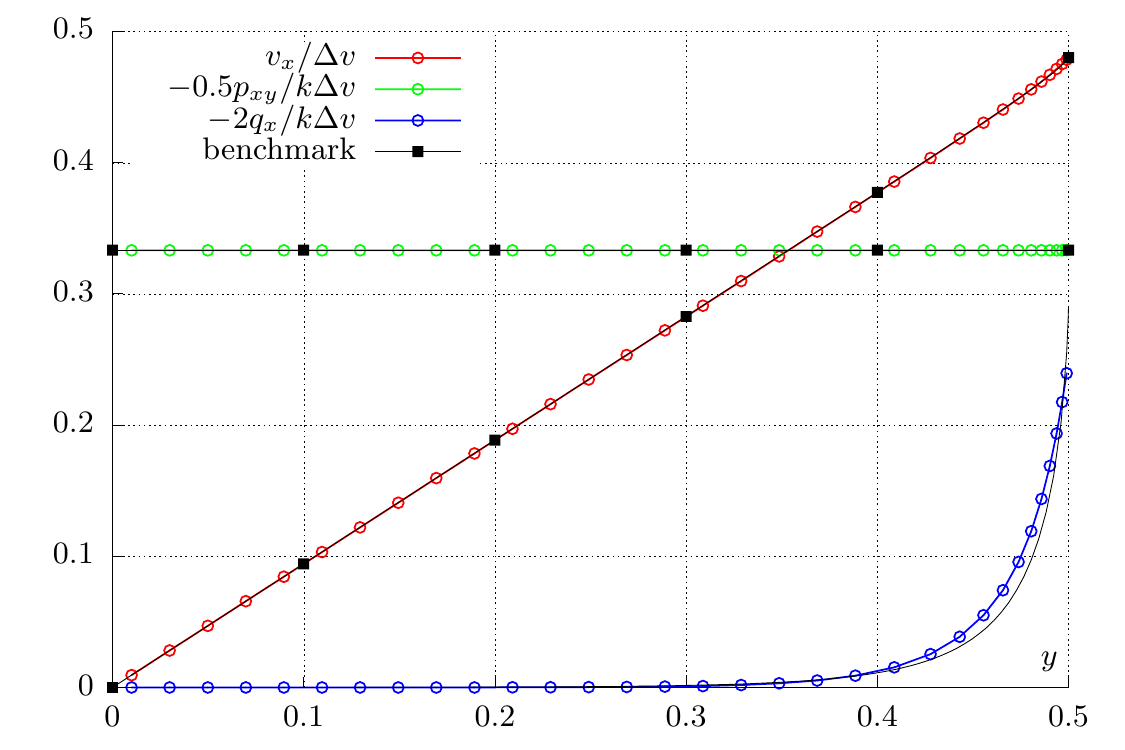}
        \caption{LB method: D3Q96}
        \label{fig:d3q96-k0_03}
    \end{subfigure}\\
    \begin{subfigure}[b]{0.5\textwidth}
        \includegraphics[width=\textwidth]{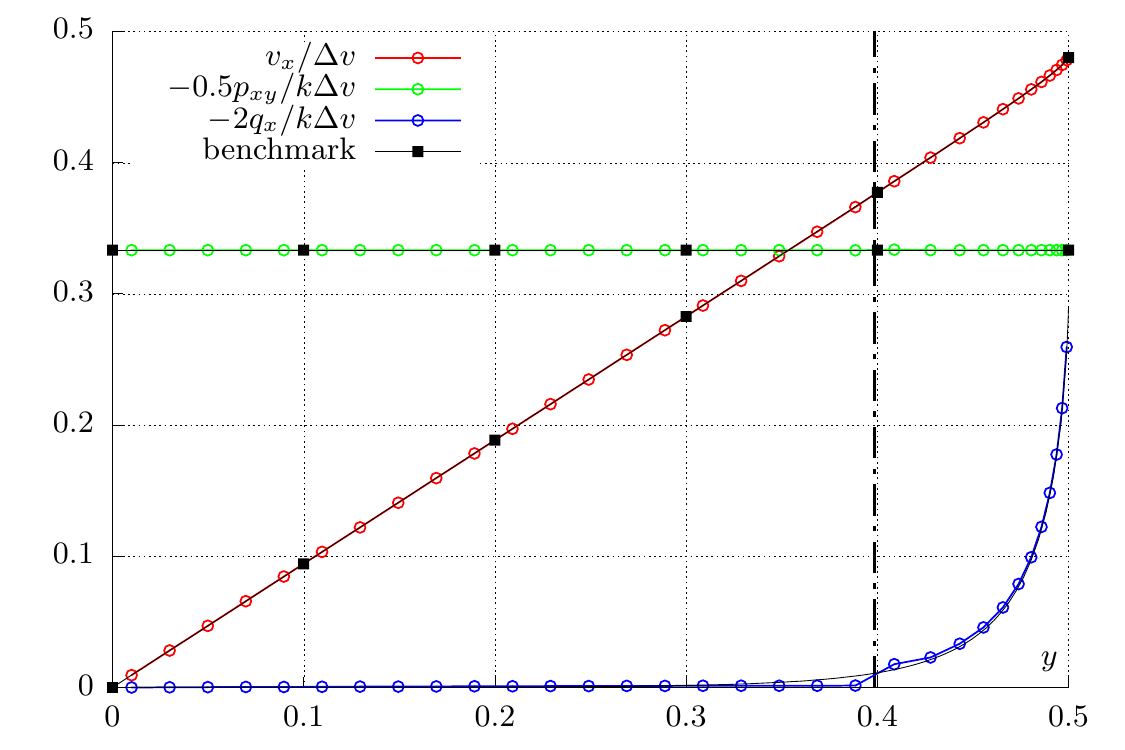}
        \caption{hybrid: DV and D3Q19}
        \label{fig:hyb:d3q19-k0_03}
    \end{subfigure}%
    \begin{subfigure}[b]{0.5\textwidth}
        \includegraphics[width=\textwidth]{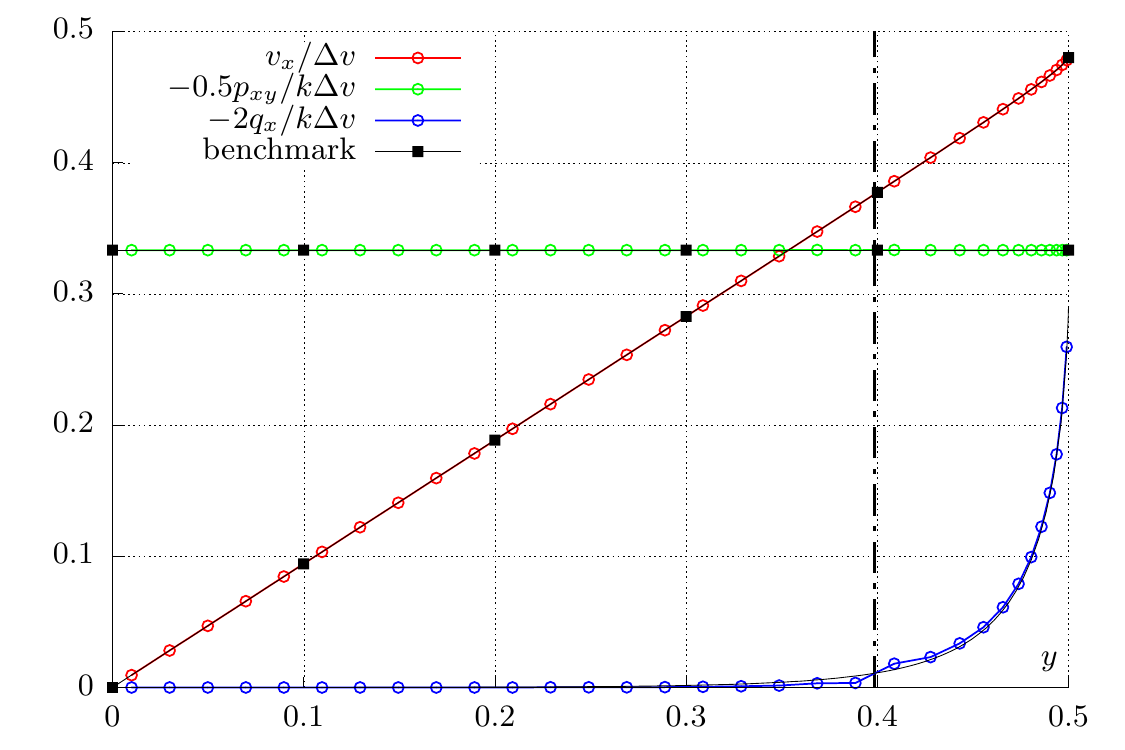}
        \caption{hybrid: DV and D3Q96}
        \label{fig:hyb:d3q96-k0_03}
    \end{subfigure}
    \caption{
        Numerical solution of the Couette-flow problem for \(k=0.03\) obtained by the proposed hybrid method.
        There are \(3\) mean free paths between the boundary and coupling interface marked with the dash-dotted line.
        The black lines are the high-accuracy benchmark solution.
        The black boxes correspond to the tabulated values from~\cite{Luo2016}.
    }\label{fig:k0_03}
\end{figure}

\begin{figure}
   \centering
   \begin{subfigure}[b]{0.5\textwidth}
       \includegraphics[width=\textwidth]{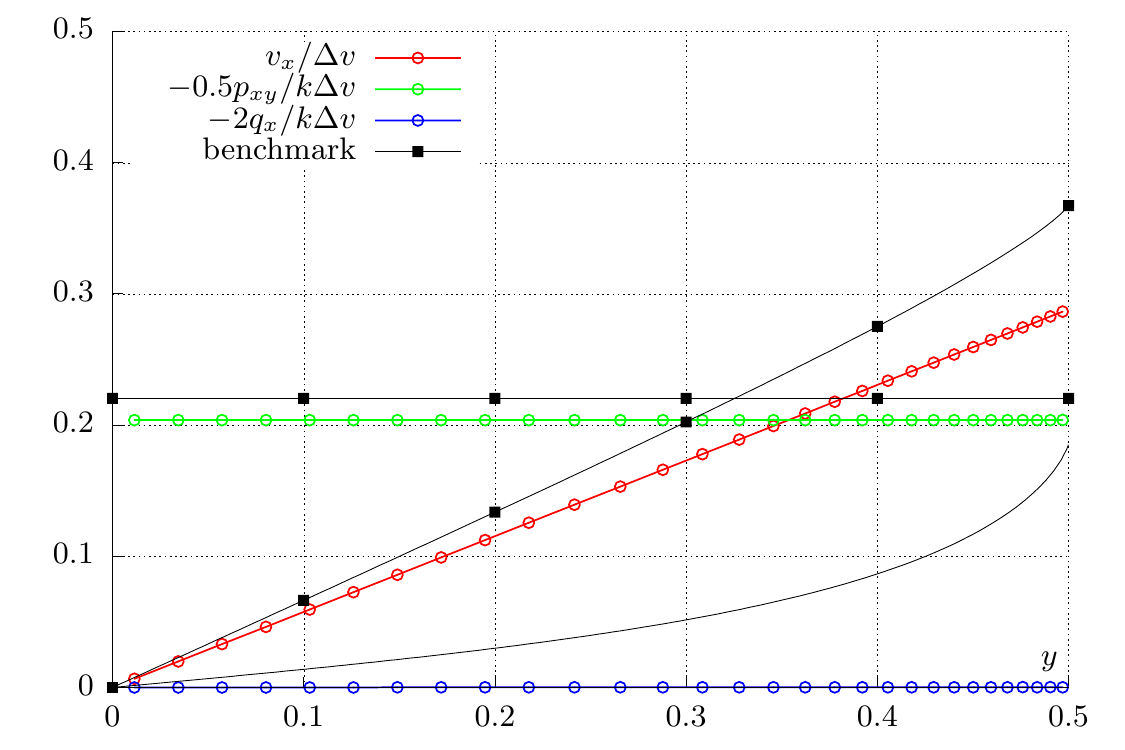}
       \caption{LB method: D3Q19}
       \label{fig:d3q19-k0_3}
   \end{subfigure}%
   \begin{subfigure}[b]{0.5\textwidth}
       \includegraphics[width=\textwidth]{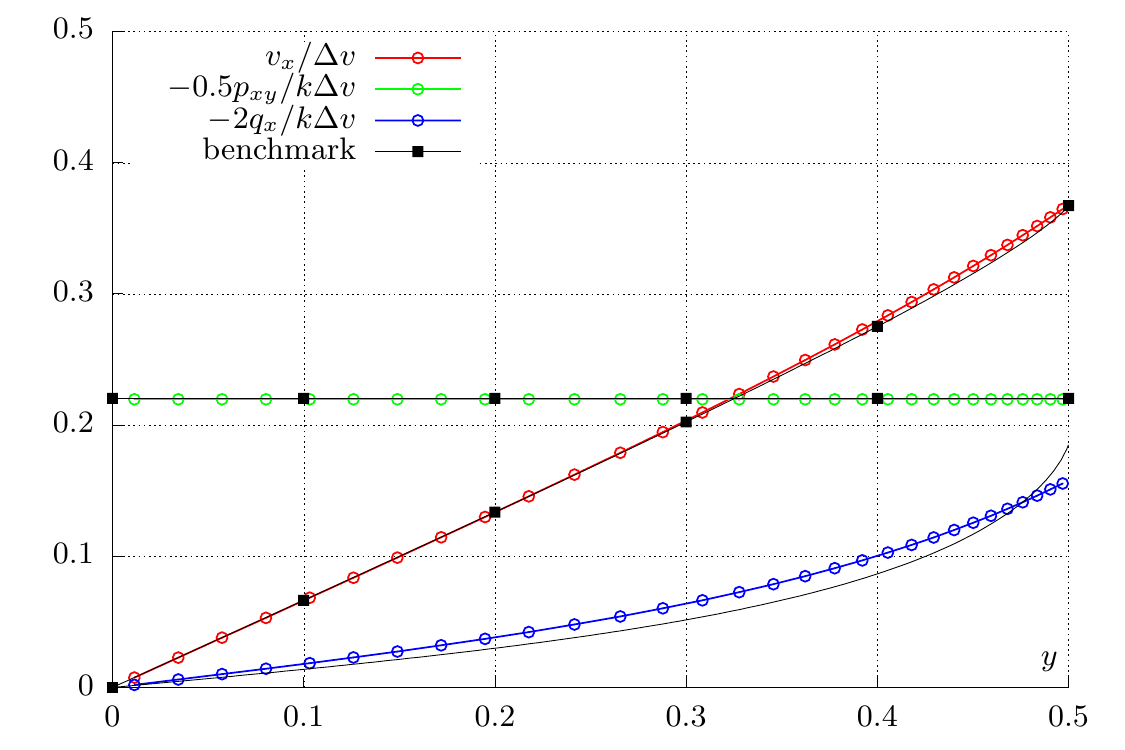}
       \caption{LB method: D3Q96}
       \label{fig:d3q96-k0_3}
   \end{subfigure}\\
   \begin{subfigure}[b]{0.5\textwidth}
       \includegraphics[width=\textwidth]{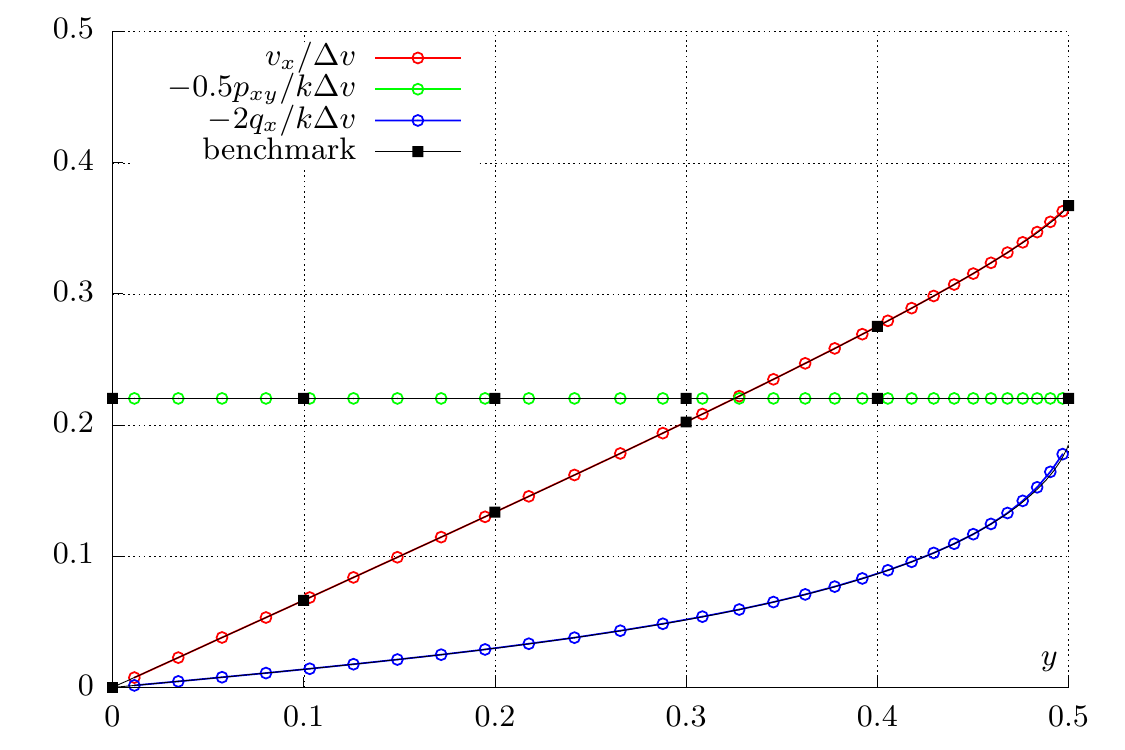}
       \caption{DV method}
       \label{fig:dvm-k0_3}
   \end{subfigure}%
   \begin{subfigure}[b]{0.5\textwidth}
       \includegraphics[width=\textwidth]{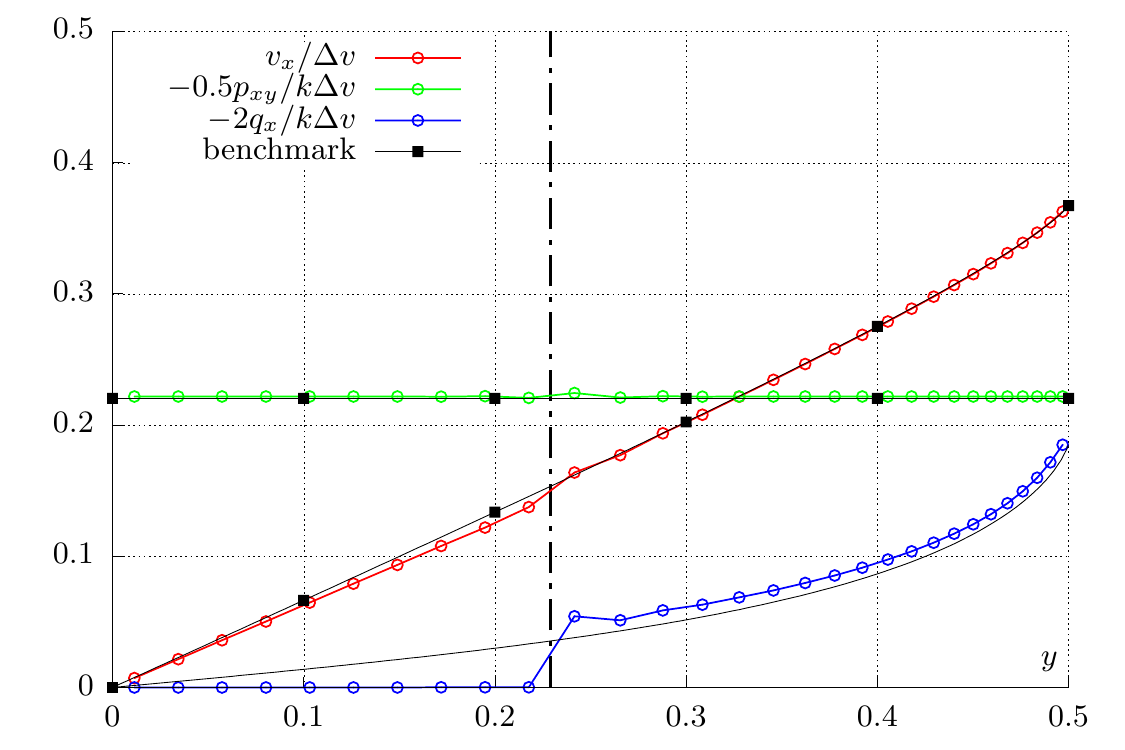}
       \caption{hybrid: DV and D3Q19}
       \label{fig:hyb:d3q19-k0_3}
   \end{subfigure}\\
   \begin{subfigure}[b]{0.5\textwidth}
       \includegraphics[width=\textwidth]{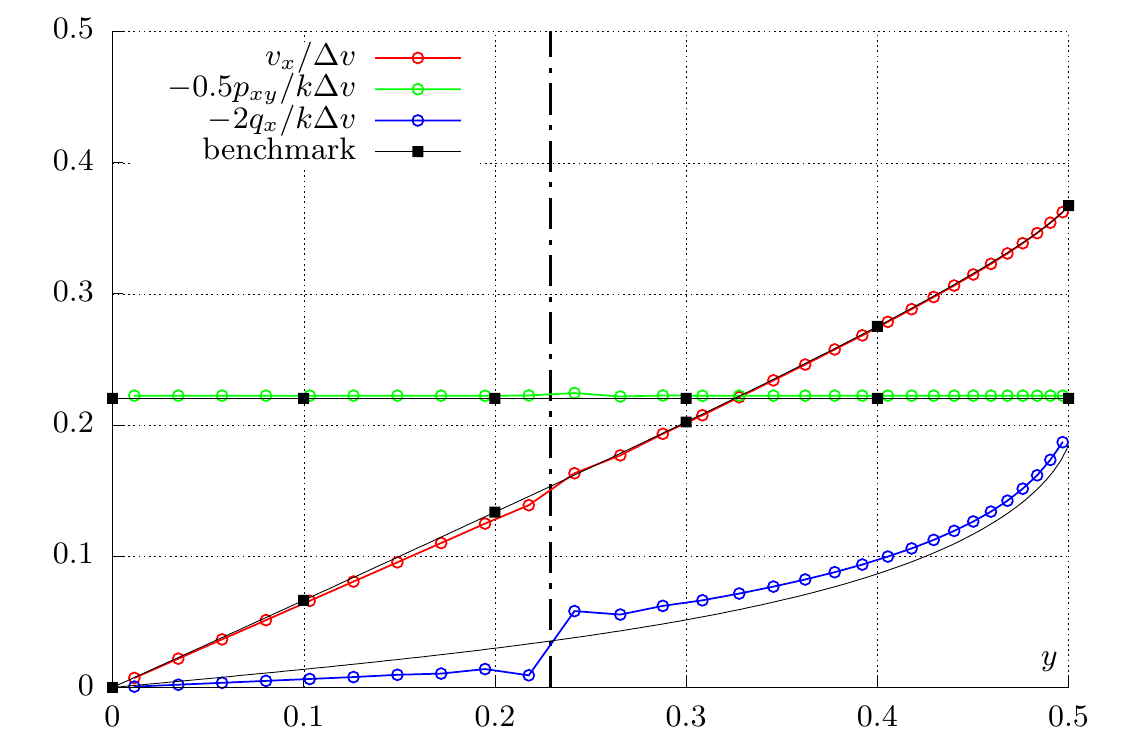}
       \caption{hybrid: DV and D3Q39}
       \label{fig:hyb:d3q39-k0_3}
   \end{subfigure}%
   \begin{subfigure}[b]{0.5\textwidth}
       \includegraphics[width=\textwidth]{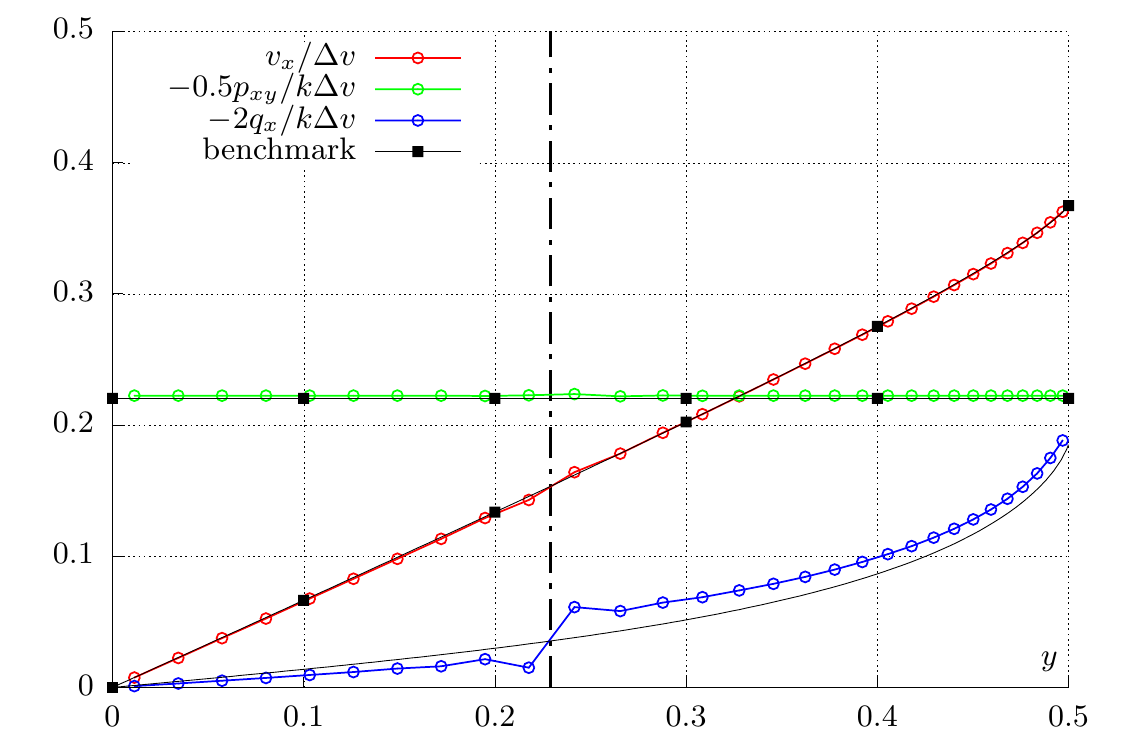}
       \caption{hybrid: DV and D3Q96}
       \label{fig:hyb:d3q96-k0_3}
   \end{subfigure}
   \caption{
       Numerical solution of the Couette-flow problem for \(k=0.3\) obtained by the proposed hybrid method.
       There are \(0.8\) mean free paths between the boundary and coupling interface marked with the dash-dotted line.
       The black lines are the high-accuracy benchmark solution.
       The black boxes correspond to the tabulated values from~\cite{Luo2016}.
   }\label{fig:k0_3}
\end{figure}

\begin{figure}
    \centering
    \begin{subfigure}[b]{0.33\textwidth}
        \includegraphics[width=\textwidth]{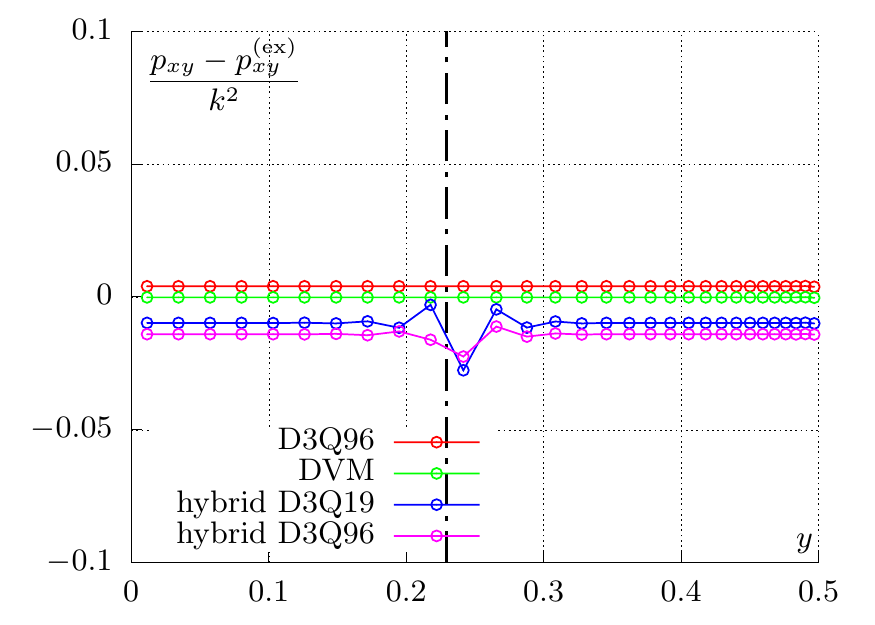}
        \caption{\(k=0.3\), \(p^\mathrm{(ex)}_{xy} = -0.1868996702\)}
        \label{fig:accuracy-k0_3}
    \end{subfigure}%
    \begin{subfigure}[b]{0.33\textwidth}
        \includegraphics[width=\textwidth]{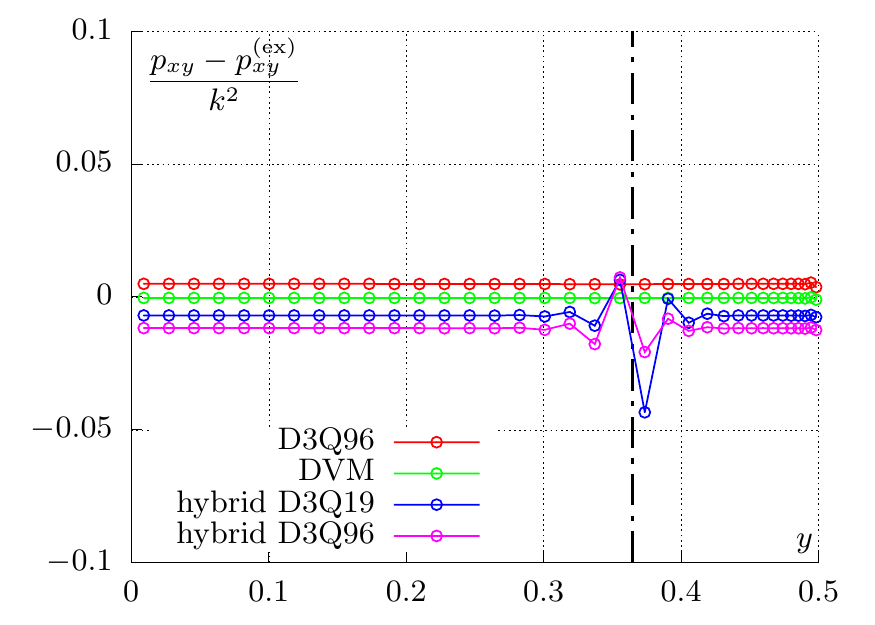}
        \caption{\(k=0.1\), \(p^\mathrm{(ex)}_{xy} = -0.08311215565\)}
        \label{fig:accuracy-k0_1}
    \end{subfigure}%
    \begin{subfigure}[b]{0.33\textwidth}
        \includegraphics[width=\textwidth]{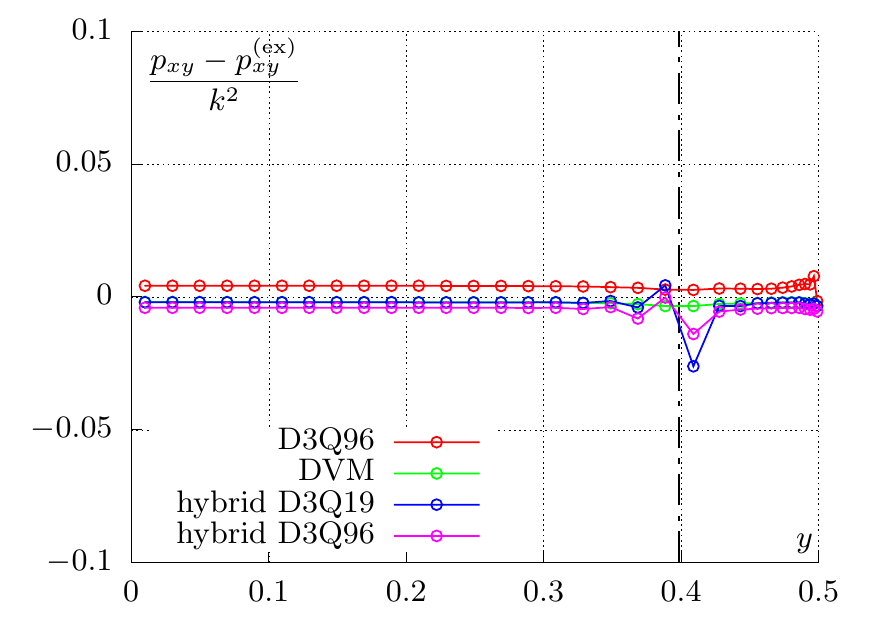}
        \caption{\(k=0.03\), \(p^\mathrm{(ex)}_{xy} = -0.02827597203\)}
        \label{fig:accuracy-k0_03}
    \end{subfigure}
    \caption{
        Absolute numerical error of the shear stress obtained by the pure and hybrid schemes.
        The dash-dotted lines correspond to the coupling interface used for the domain decomposition.
        The reference values \(p^\mathrm{(ex)}_{xy}\) are taken from~\cite{Luo2016}.
    }\label{fig:accuracy}
\end{figure}

%%% Pure DV and LB results
The numerical results obtained by the pure DV and LB methods for \(k=0.1\) are shown in Fig.~\ref{fig:pure}.
The nonuniform velocity grid refined at the sharp variations of the VDF
yields a small discrepancy between the DV and benchmark profiles (Fig.~\ref{fig:dvm}).
As for the LB method, the 5-order D3Q19 model and 7-order D3Q39 model~\cite{Shan2006}
based on the Gauss--Hermite quadratures are considered, along with special 7-order D3Q96 model
developed for the boundary-value problems driven by the diffuse-reflection boundary condition~\cite{Feuchter2016}.
Ability to capture rarefied-gas effects arising from the kinetic boundary conditions
is observed from the profile of the longitudinal heat flux \(q_x\).
In particular, models of the Navier--Stokes level do not capture it due to lack of additional degrees of freedom,
e.g., the D3Q19 model does not cover the third-order moments of the VDF (Fig.~\ref{fig:d3q19}).
Instead, there is a small spurious positive heat flux in Fig.~\ref{fig:d3q19},
which is \(\OO{\Delta v^3}\) and closely associated with the stress tensor and velocity.
The model D3Q39 partially reproduces the heat flux (Fig.~\ref{fig:d3q39}),
while the D3Q96 \(q_x\) profile appears to be quite close to the exact one (Fig.~\ref{fig:d3q96}).

%%% Why D3Q96 is better than D3Q39
Increasing the order of the LB model helps to capture the corresponding low-order moments of the VDF
but failed to describe its high-order relaxation correctly.
However, the LB models augmented by special velocity groups are capable of reproducing the Knudsen layer to some extent.
The augmented model D3Q96 reproduces the Maxwell half-range (or wall) moments better than D3Q39,
which results in better reproduction of the diffuse-reflection boundary condition.
As a result, D3Q96 produces qualitatively correct results for the longitudinal heat flux,
which has non-thermo-hydrodynamical nature, although it does not recover thermo-hydrodynamics.

%%% Hybrid results
The numerical results for the hybrid schemes are shown in Fig.~\ref{fig:hybrid}.
Quantities \(v_x\) and \(p_{xy}\) are close to the exact solution,
but there is a noticeable distortion behind the coupling interface in the D3Q19 velocity profile.
Hybrid \(q_x\) is close to the pure DV one only in the kinetic region (the DV part of the hybrid solution).
There are small oscillations of macroscopic variables in the buffer zone,
and they are particularly noticeable for \(q_x\), since its profile is multiplied by factor 40.
The amplitude of these oscillations is proportional to the high-order terms of the Hermite expansion of the VDF
that are not included in the employed mapping method.
These terms decrease exponentially as the coupling interface moves away from \(y=1/2\).
The numerical results for \(k=0.03\) shown in Fig.~\ref{fig:k0_03},
where 3 mean free paths (in contrast to 1.2 for \(k=0.1\)) are covered by DV method, clearly illustrate this fact.

%%% Case k=0.3
For larger \(k\), the advantage of the DV method over the LB one in terms of accuracy becomes more evident,
since rarefied-gas effects amplify considerably (Fig.~\ref{fig:k0_3}).
The D3Q19 model loses its accuracy most of all:
there is a substantial deviation from the benchmark profiles of \(p_{xy}\) and \(v_x\) in Fig.~\ref{fig:d3q19-k0_3}.
Although D3Q96 can reproduce half-range (wall) moments most accurately within the given order,
it fails to reproduce the complex structure of a highly nonequilibrium VDF with sharp variations
around plane \(\xi_y=0\) (Fig.~\ref{fig:d3q96-k0_3}).
In contrast, the DV method with the velocity grid refined near \(\xi_y=0\) adequately copes with this task.
One cannot conclude the same for the results obtained by hybrid schemes.
There is a noticeable discrepancy for \(q_x\) and a kink for \(v_x\) in Fig.~\ref{fig:hyb:d3q19-k0_3}--\subref{fig:hyb:d3q96-k0_3}.
The proximity of the coupling interface to the boundary in terms of mean free path is the primary reason
for the observed lack of accuracy. Note also that the hybrid method based on the D3Q96 model has a minimal kinking
of the velocity profile in comparison to D3Q19 and D3Q39.

%%% Accuracy of results
The shear stress profiles look constant in Figs.~\ref{fig:hybrid},~\ref{fig:k0_03}, and~\ref{fig:k0_3},
since the absolute error is everywhere smaller than \(0.06k^2\), which is easily seen in Fig.~\ref{fig:accuracy}.
The largest error is observed in the points in the vicinity of the coupling interface.
In the absence of conservative correction of fluxes on the coupling interface~\eqref{eq:poly_correction},
these oscillations turn into monotonic jumps, which introduce much more error into results.
This fact clearly confirms the well-known importance of preserving the conservation properties by a numerical method.
Incidentally, let us note that the obtained numerical accuracy is sufficient
to distinguish molecular potentials~\cite{Sharipov2013, Su2018}.

%%% [Aristov]: We compared our results with available data on the benchmark problem [Phys. Fluids 25, 027101 (2013)] devoted to the computation ab initio by DSMC for the Couette flow problem. In [1] for U/u0=0.2 the values for comparison of only the shear stress are given (the temperature and the velocity gradient are not presented). These results ab initio are compared with the analogous calculations for HS and from Table IV in [1] one can see that the relative difference is of 0.3% (the column concerning C=0 or C=1). Our value of the shear stress for the hybrid scheme is 0.83 and differs from the tabulated values for the Couette flow is of the same order, namely 0.3%. So we conclude that the difference between the hybrid shear stress results and the value of the benchmark results ab initio is in the range of the computational error.

\begin{figure}
    \centering
    \includegraphics[width=0.6\textwidth]{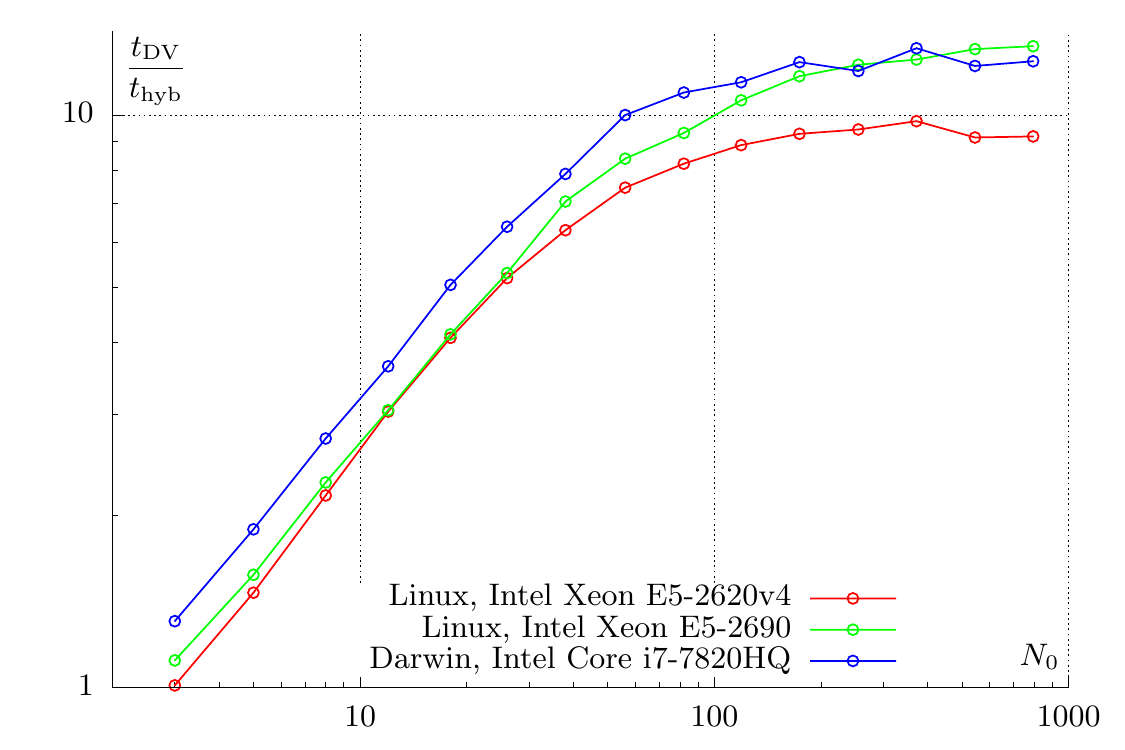}
    \caption{
        Computational speed-up yielded by the hybrid method for different CPUs and operational systems.
        $N_0$ is the number of cells in the kinetic zone, $10N_0$ is the number of cells in the bulk region,
        $t_\DV$ and $t_\mathrm{hyb}$ are the total CPU times elapsed by the DV and hybrid methods, respectively.
    }\label{fig:speed-up}
\end{figure}

%%% Efficiency
Finally, let us touch upon the efficiency of the proposed hybrid scheme.
The computational speed-up with respect to the pure DV scheme is shown in Fig.~\ref{fig:speed-up}
as a ratio of the corresponding CPU times,
while the ratio of cells in the kinetic and bulk regions remains constant.
One can see that the efficiency of the hybrid method achieves the optimum value
when the number of cells in the kinetic region is more than $10^2$.
Note that the asymptotic speed-up can be slightly higher than
the optimum one (12--13 versus 11 in Fig.~\ref{fig:speed-up}).
It is mainly due to memory saving, which results in fewer cache misses.

\subsection{Breakdown criterion}

\begin{figure}
    \centering
    \begin{subfigure}[b]{0.5\textwidth}
        \includegraphics[width=\textwidth]{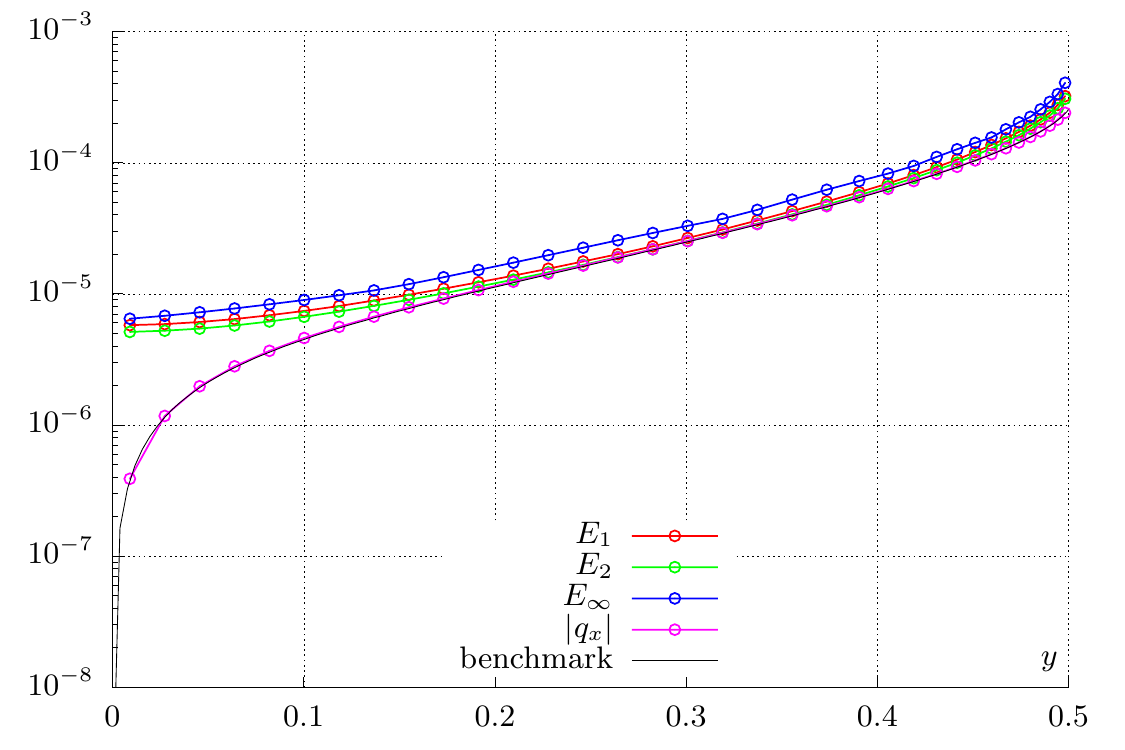}
        \caption{DV method}
        \label{fig:norms:dvm}
    \end{subfigure}%
    \begin{subfigure}[b]{0.5\textwidth}
        \includegraphics[width=\textwidth]{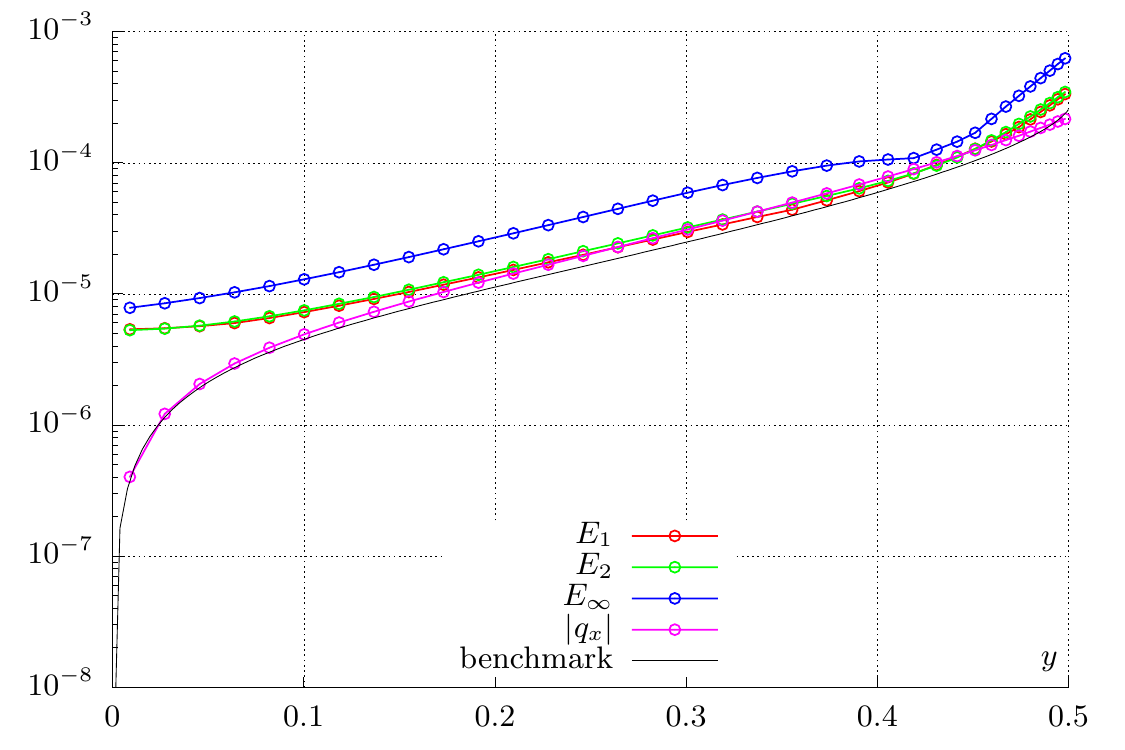}
        \caption{LB method: D3Q96}
        \label{fig:norms:d3q96}
    \end{subfigure}\\
    \begin{subfigure}[b]{0.5\textwidth}
        \includegraphics[width=\textwidth]{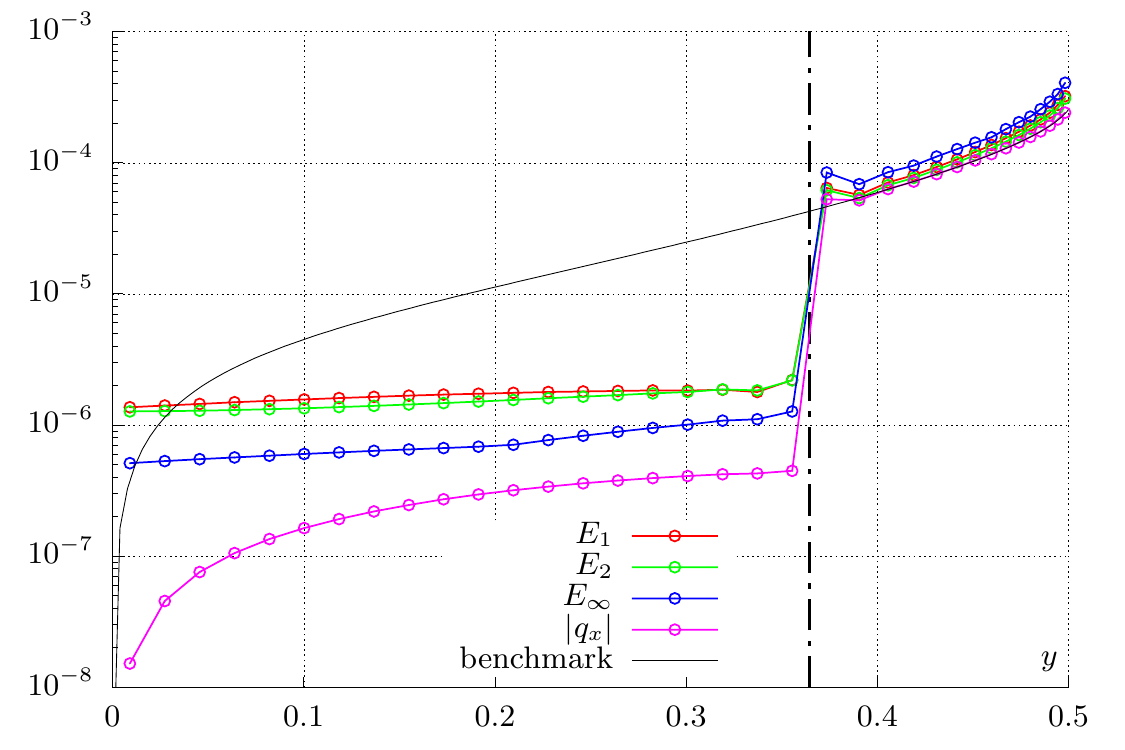}
        \caption{hybrid: DV and D3Q19}
        \label{fig:norms:d3q19-hyb}
    \end{subfigure}%
    \begin{subfigure}[b]{0.5\textwidth}
        \includegraphics[width=\textwidth]{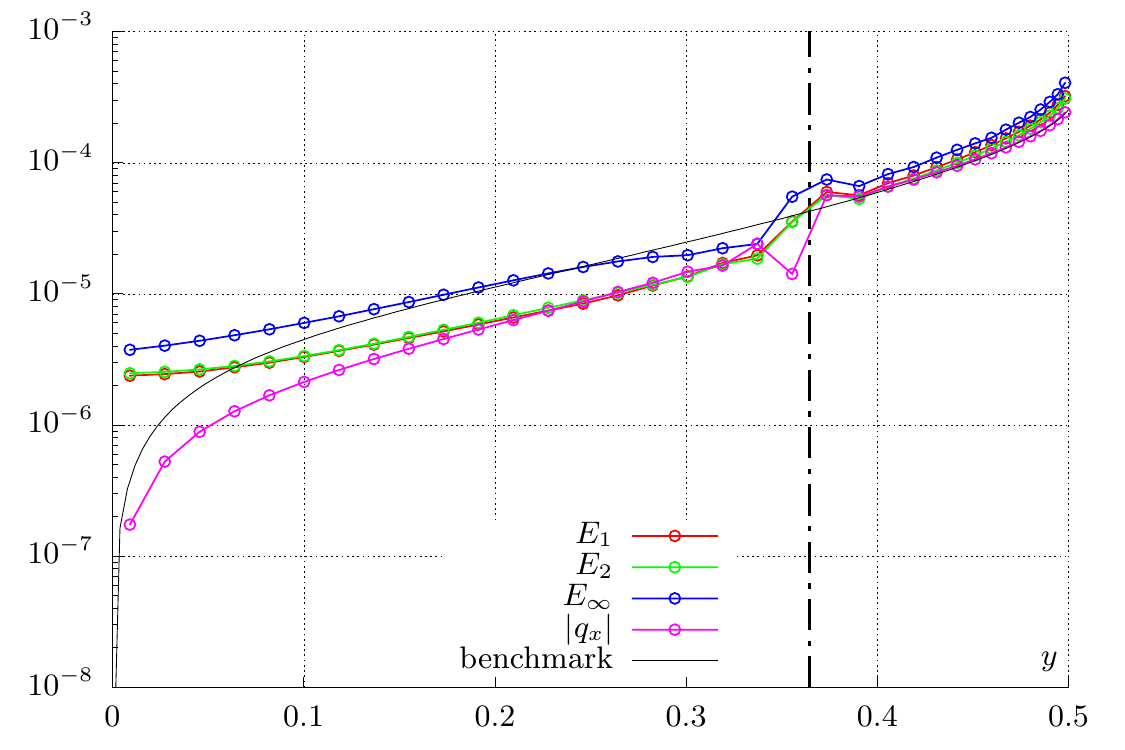}
        \caption{hybrid: DV and D3Q96}
        \label{fig:norms:d3q96-hyb}
    \end{subfigure}
    \caption{
       Quantities that can serve as a equilibrium breakdown parameter for \(k=0.1\).
    }\label{fig:norms}
\end{figure}

%%% Breakdown criterion and adaptive algorithm
A multiscale hybrid method based on the domain decomposition procedure should be supplied
with the so-called equilibrium breakdown criterion, which determines the position of the coupling interface
between different models for approximation of the VDF. The typical way to quantify this criterion is
to introduce an appropriate equilibrium breakdown parameter \(E\) and compare it with some predefined value \(\epsilon_E\).
The DV method is used only in nonequilibrium regions, specifically in cells, where \(E>\epsilon_E\),
while the LB method is employed in the others.

%%% Breakdown parameters
Longitudinal heat flux \(q_x\) appears only in the Knudsen layer and, therefore, can serve as an equilibrium breakdown parameter
for the investigated Couette-flow problem, but not in the general case.
Criteria based on deviation of the VDF from the truncated Chapman--Enskog expansion is natural for kinetic schemes.
For instance, quantities \(E_p=\|f-\NSF{f}\|_p/\|f\|_p\),
the deviation from the Navier--Stokes--Fourier (NSF) order of approximation~\cite{Zhang2014}:
\begin{gather}
    \NSF{f}_{\DV,j} = \equil{f}_{\DV,j}\br{
        1 + \frac{\caj\cbj P_{\alpha\beta}}{2pT} + \frac{\caj q_\alpha}{pT}\br{ \frac{c_j^2}{5T} - 1 } }, \label{eq:NSF_DV} \\
    \NSF{f}_{\LB,j} = \equil{f}_{\LB,j} + w_j\xiaj\br{
         P_{\alpha\beta}\br{\xibj\br{1 + \xigj v_\gamma} - 2 v_\beta } + q_\alpha\br{ \frac{\xi_j^2}5 - 1 } }, \label{eq:NSF_LB}
\end{gather}
where \(P_{\alpha\beta} = p_{\alpha\beta} - \rho T\delta_{\alpha\beta}\) and \(\caj = \xiaj - v_\alpha\),
are shown in Fig.~\ref{fig:norms} for the following norms in the discrete velocity space:
\begin{equation}\label{eq:norms}
    \|f\|_p = \bigg(\sum_j |f_j|^p \bigg)^{1/p}, \quad p=1,2, \quad \|f\|_\infty = \max_j |f_j|.
\end{equation}

%%% Discussion on figures
The D3Q19 model produces an almost constant profile (Fig.~\ref{fig:norms:d3q19-hyb})
since it describes nothing beyond the NSF level.
The D3Q96 profile (Fig.~\ref{fig:norms:d3q96}) is close to the DV one (Fig.~\ref{fig:norms:dvm}),
which indirectly indicates that this LB model gives an acceptable approximation for the Couette-flow problem.
Due to the diffuse-reflection boundary condition, there is a discontinuity of the VDF on the boundary,
which decays monotonically and faster than any inverse power of distance from the boundary.
Therefore, all the breakdown parameters reach their maximum on the boundary;
however, \(E_\infty\) relaxes in a non-smooth way.
It is probably due to crude approximation of the sharp variations of the VDF in the Knudsen layer.
For the D3Q96 model, \(E_\infty\) noticeably exceeds \(E_{1,2}\) (Fig.~\ref{fig:norms:d3q96}),
which can be explained by its peculiar properties minimizing the wall moment errors.
The Hermite-based coupling induces oscillations (Fig.~\ref{fig:norms:d3q19-hyb},~\ref{fig:norms:d3q96-hyb})
since it is unable to reconstruct nonequilibrium part of the VDF.
The sharp drop in Fig.~\ref{fig:norms:d3q19-hyb} indicates that the coupling interface is too close to the boundary,
while the smoother transition in Fig.~\ref{fig:norms:d3q96-hyb} can be considered as more acceptable.
Nevertheless, both hybrid schemes have almost the same DV part of the solution.

\section{Conclusions and perspectives}\label{sec:summary}

%%% What have we done?
In this paper, we have presented a new algorithm for coupling the LB and DV methods
for solving the Boltzmann kinetic equation.
The Couette flow of a rarefied BGK gas has been analyzed numerically as a test example.
The continuum gas behavior is described by several Gauss--Hermite LB models
with various numbers of discrete velocities ranging from 19 to 96.
Incorporating the augmented~\cite{Feuchter2016} LB models
positively affects the solution accuracy in comparison to the conventional LB models.
The Knudsen layer is captured accurately by the DV method with highly nonuniform velocity grids.
The numerical stress-tensor anisotropy generated by such grids is eliminated
by minimizing the 7-moment discrete entropy functional.
The physical mesh refinement near the boundary allows to effectively approximate the weak singularity
of the profiles of the macroscopic variables.
The LB and DV solutions are matched by means of projection onto the truncated Hermite basis.
The second-order FV solution of both subdomains are supplied by two ghost cells with the reconstructed values.
The additional polynomial correction procedure of the interfacial DV flux has been employed
to ensure conservative properties of the coupling algorithm.

%%% Applicability and comparison with others
Since the employed LB models are suited for the isothermal hydrodynamics,
the proposed hybrid method is applicable to flows with Mach number $\lesssim0.3$.
The subsonic regime is quite challenging for the DSMC-based methods due to the inherent stochastic noise,
but the deterministic approaches work well.

%%% Other LB models
The other LB models (e.g., for supersonic flows, compressible, and thermal flows~\cite{Chen2010, Frapolli2015, Frapolli2016})
can be potentially incorporated into the hybrid method.
The entropic models~\cite{Karlin1999, Chikatamarla2006, Chikatamarla2009} are promising
due to their enhanced stability for low viscosities (large Reynolds numbers).
Using of the regularized high-order LB models~\cite{Zhang2006, Mont2015, Mattila2017} for the hybrid schemes is interested as well.
One can conclude that the improvements in employed LB model should be directed
to the better reproduction of the higher Maxwell moments and half-range (wall) moments~\cite{Ambrus2016}.
In this case, the mapping method should also be upgraded by involving higher-order and wall moments.

%%% Adaptive mesh in velocity space
Adaptive coupling of the LB and DV methods under the unified FV formulation can be considered
as a variant of the DV methodologies based on adaptive grids in velocity space~\cite{Aristov1977, Kolobov2013, Baranger2014}.
The adaptation of the DV set according to the local flow regime provides room for improving the efficiency of numerical methods
and can serve as a foundation of hybrid schemes for multiscale compressible flows.

%%% Ilyin: applicability of our method
% The present scheme can be extended to non-negligible $\Ma$ flows.
% Since the applied high-order LB model reproduces isothermal hydrodynamics exactly,
% the method can be applicable to the flows with $\Ma \sim 0.1-0.3$.

% We emphasize that LB is less computationally demanding than DV method.
%On the other hand, conventional LB is aimed to reproduce only the lowest moments of VDF in slow regimes
%whereas DV is able to cope with strong nonequilibrium effects.

\section*{Acknowledgements}

This work was supported by the Russian Foundation for Basic Research (Grants 18-01-00899, 18-07-01500).
The authors are grateful to the referees for many constructive comments that helped in improving the quality of the paper.

\appendix

\section{Third-order TVD limiter for the FV scheme}\label{sec:limiter}

Let us approximate the one-dimensional transport equation \(\Pder[f]{t} + \xi\Pder[f]{x}\)
in the FV manner with constant mesh (\(\Delta{x}\)) and time (\(\Delta{t}\)) steps:
\begin{equation}\label{eq:fv:scheme}
    f^{n+1}_m = f^n_m - \gamma\br{f^{n+1/2}_{m+1/2} - f^{n+1/2}_{m-1/2}}, \quad
    \gamma = \frac{\xi\Delta{t}}{\Delta{x}} \geq 0, \quad
    m = 1,\dots,M, \quad n\in\mathbb{N},
\end{equation}
where the reconstructed edge values for the third-point stencil are written in the following form:
\begin{equation}\label{eq:fv:flux}
    f^{n+1/2}_{m+1/2} = f^n_m + \frac{1-\gamma}2\phi\br{\frac{\Delta_-}{\Delta_+}}\Delta_+, \quad m = 1,\dots,M,
\end{equation}
where \(\Delta_\pm = \pm\br{f^n_{m\pm1} - f^n_m}\), function \(\phi(\theta)\) is usually called a flux limiter.
Within the one-point stencil for positive velocities \(\xi_j>0\),
it is possible to construct the first-order (upwind) scheme only (\(\phi=0\)).
The second-order accuracy can be achieved, if \(\phi(1+x) = 1 + \OO{x}\).
For the third-order accuracy, the following condition is sufficient:
\begin{equation}\label{eq:fv:third_order}
    \phi(1+x) = 1 + \frac{1+\gamma}3x + \OO{x^2}.
\end{equation}
Taking into account the CFL condition \(\gamma\leq1\)
and non-negativity of the limiter for monotonic functions \(\phi(\theta\geq0)\geq0\),
one can obtain the following sufficient conditions for the TVD property of the numerical scheme:
\begin{equation}\label{eq:fv:tvd}
    \begin{cases}
        \phi(\theta) = 0, \quad \theta < 0, \\
        \phi(\theta) \leq \dfrac{2}{\gamma}\theta, \quad \theta \geq 0, \\
        0 \leq \phi(\theta) \leq \dfrac2{1-\gamma}.
    \end{cases}
\end{equation}
Combining~\eqref{eq:fv:third_order} and~\eqref{eq:fv:tvd}, one can construct a third-order TVD limiter:
\begin{equation}\label{eq:fv:limiter}
    \phi(\theta) = \begin{cases}
        0, &\quad \theta < 0, \\
        \min\br{ \dfrac{2}{\gamma}\theta, \dfrac{1+\gamma}3\theta + \dfrac{2-\gamma}3, \dfrac2{1-\gamma}},
            &\quad \theta \geq 0,
    \end{cases}
\end{equation}
which is used in the present study, but for nonuniform mesh step~\eqref{eq:limiter}.
Let us note that inequalities~\eqref{eq:fv:tvd} are often replaced by stronger ones, independent of \(\gamma\),
especially when nonlinear transport equation is considered.

\section{Solution of the linear Couette-flow problem for the BGK model}\label{sec:benchmark_solution}

%%% Formulation of the linear Couette-flow problem
For symmetry reasons, the steady-state behavior of the gas in the Couette flow at \(\Delta{v}\to0\)
and arbitrary \(k\) is described by the following VDF:
\begin{equation}\label{eq:exact:vdf}
    f(\bxi) = \omega(\bxi)\br{ 1 + \Delta{v}\xi_x \Phi(y, \xi_y, \xi) }, \quad \xi = |\bxi|,
\end{equation}
where \(\omega(\bxi)\) is defined in~\eqref{eq:omega} and \(\Phi\) is governed by
\begin{equation}\label{eq:exact:Phi}
    \tau\xi_y\pder[\Phi]{y} = \frac{v_x}{\Delta{v}} - \Phi, \quad\
    \Phi\br{ y=\mp\frac12, \xi_y\gtrless0, \xi } = \mp\frac12,
\end{equation}
which is derived under the assumption of the BGK model of collisional term~\cite{Sone2007}.
The macroscopic variables are the following moments of \(\Phi\):
\begin{equation}\label{eq:exact:macro}
    v_x = \Delta{v}\int \xi_x^2\Phi\omega\dd\xi, \quad
    p_{xy} = \Delta{v}\int \xi_x^2\xi_y\Phi\omega\dd\xi, \quad
    q_x = \frac{\Delta{v}}2 \int \xi_x^2\xi^2\Phi\omega\dd\xi - \frac52v_x.
\end{equation}

%%% Solution
The solution of~\eqref{eq:exact:Phi} can be represented as
\begin{equation}\label{eq:exact:Phi_explicit}
    \Phi\br{ |y| \leq \frac12, \xi_y\gtrless0, \xi } = \mp\frac12\exp\br{ \mp\frac{1\pm2y}{2\tau\xi_y} } +
        \frac1{2\tau\xi_y} \int_{\mp\frac12}^y \exp\br{ -\frac{y-s}{\tau\xi_y} } g(s)\dd{s} ,
\end{equation}
where \(g(y) = 2v_x/\Delta v\) is obtained from the following integral equation~\cite{Willis1962}:
\begin{equation}\label{eq:exact:g}
    \sqrt{\pi} g(y) = \mathcal{J}_0 \left(\frac{1-2y}{2k}\right) - \mathcal{J}_0 \left(\frac{1+2y}{2k}\right)
        + \frac1k \int_0^{\frac12} \left[ \mathcal{J}_{-1}\left(\frac{|y-s|}{k}\right)
        - \mathcal{J}_{-1}\left(\frac{y+s}{k}\right) \right] g(s) \dd{s}.
\end{equation}
Here, \(\mathcal{J}_n(s)\) are the Abramowitz functions~\cite{Abramowitz1972}:
\begin{equation}\label{eq:exact:Abramowitz}
    \mathcal{J}_n(s) = \int_0^\infty t^n \exp\br{ -t^2-\frac{s}{t} } \dd t,
    \quad s\ge0, \quad n \in \mathbb{Z}.
\end{equation}
The remaining macroscopic variables are calculated as follows:
\begin{gather}
    \frac{p_{xy}}{\Delta v} = -\frac{\tau}{\sqrt{\pi}} \br{
        \mathcal{J}_2(0)-\mathcal{J}_2\left(\frac1k\right)
        + \frac1k\int_0^{\frac12}\left[
            \mathcal{J}_1\left(\frac{1-2s}{2k}\right)-\mathcal{J}_1\left(\frac{1+2s}{2k}\right)
        \right]g(s)\dd{s}
    }, \label{eq:exact:Pxy} \\
    \begin{aligned}
    \frac{q_x}{\Delta v} = \frac1{2\sqrt{\pi}} & \left(\vphantom{\int_0^{\frac12}}
        \mathcal{J}_2\br{ \frac{1-2y}{2k} } - \mathcal{J}_2\br{ \frac{1+2y}{2k} }\right. \\
        &+\left. \frac1k\int_0^{\frac12}\left[
            \mathcal{J}_1\br{ \frac{|y-s|}k } - \mathcal{J}_1\br{ \frac{y+s}k }
        \right]g(s)\dd{s}
    \right) - \frac{g(y)}4.
    \end{aligned} \label{eq:exact:qflow}
\end{gather}
Methods of a high-accuracy solution of~\eqref{eq:exact:g} are presented in~\cite{Luo2015, Luo2016}.

\bibliography{dvm-lbm}

\begin{thebibliography}{10}
\expandafter\ifx\csname url\endcsname\relax
  \def\url#1{\texttt{#1}}\fi
\expandafter\ifx\csname urlprefix\endcsname\relax\def\urlprefix{URL }\fi
\expandafter\ifx\csname href\endcsname\relax
  \def\href#1#2{#2} \def\path#1{#1}\fi

\bibitem{Dimarco2014}
G.~Dimarco, L.~Pareschi, Numerical methods for kinetic equations, Acta Numer.
  23 (2014) 369--520.
\newblock \href {https://doi.org/10.1017/S0962492914000063}
  {\path{doi:10.1017/S0962492914000063}}.

\bibitem{Bourgat1996}
J.-F. Bourgat, P.~Le~Tallec, M.~Tidriri, Coupling boltzmann and navier--stokes
  equations by friction, J. Comput. Phys. 127~(2) (1996) 227--245.
\newblock \href {https://doi.org/10.1006/jcph.1996.0172}
  {\path{doi:10.1006/jcph.1996.0172}}.

\bibitem{Tallec1997}
P.~Le~Tallec, F.~Mallinger, Coupling boltzmann and navier--stokes equations by
  half fluxes, J. Comput. Phys. 136~(1) (1997) 51--67.
\newblock \href {https://doi.org/10.1006/jcph.1997.5729}
  {\path{doi:10.1006/jcph.1997.5729}}.

\bibitem{Potkin1975}
V.~V. Potkin, Kinetic analysis of difference schemes for gas dynamics, USSR
  Comp. Math. Math. Phys. 15~(6) (1975) 126--132.
\newblock \href {https://doi.org/10.1016/0041-5553(75)90208-6}
  {\path{doi:10.1016/0041-5553(75)90208-6}}.

\bibitem{Pullin1980}
D.~I. Pullin, Direct simulation methods for compressible inviscid ideal-gas
  flow, J. Comput. Phys. 34~(2) (1980) 231--244.
\newblock \href {https://doi.org/10.1016/0021-9991(80)90107-2}
  {\path{doi:10.1016/0021-9991(80)90107-2}}.

\bibitem{Reitz1981}
R.~D. Reitz, One-dimensional compressible gas dynamics calculations using the
  boltzmann equation, J. Comput. Phys. 42~(1) (1981) 108--123.
\newblock \href {https://doi.org/10.1016/0021-9991(81)90235-7}
  {\path{doi:10.1016/0021-9991(81)90235-7}}.

\bibitem{Aristov1983}
V.~V. Aristov, F.~G. Cheremisin, A solution to euler and navier-stokes
  equations based on the operator splitting of a kinetic equation, Dokl. Akad.
  Nauk SSSR+ 272~(3) (1983) 555--559.

\bibitem{Elizarova1985}
T.~G. Elizarova, B.~N. Chetverushkin, Kinetic algorithms for calculating gas
  dynamic flows, USSR Comp. Math. Math. Phys. 25~(5) (1985) 164--169.
\newblock \href {https://doi.org/10.1016/0041-5553(85)90194-6}
  {\path{doi:10.1016/0041-5553(85)90194-6}}.

\bibitem{Deshpande1986}
S.~M. Deshpande, Kinetic theory based new upwind methods for inviscid
  compressible flows, in: 24th Aerospace Sciences Meeting, 1986, p. 275.
\newblock \href {https://doi.org/10.2514/6.1986-275}
  {\path{doi:10.2514/6.1986-275}}.

\bibitem{Prendergast1993}
K.~H. Prendergast, K.~Xu, Numerical hydrodynamics from gas-kinetic theory, J.
  Comput. Phys. 109~(1) (1993) 53--66.
\newblock \href {https://doi.org/10.1006/jcph.1993.1198}
  {\path{doi:10.1006/jcph.1993.1198}}.

\bibitem{Chou1997}
S.-Y. Chou, D.~Baganoff, Kinetic flux--vector splitting for the navier--stokes
  equations, J. Comput. Phys. 130~(2) (1997) 217--230.
\newblock \href {https://doi.org/10.1006/jcph.1996.5579}
  {\path{doi:10.1006/jcph.1996.5579}}.

\bibitem{Ohwada2004Xu}
T.~Ohwada, K.~Xu, The kinetic scheme for the full-burnett equations, J. Comput.
  Phys. 201~(1) (2004) 315--332.
\newblock \href {https://doi.org/10.1016/j.jcp.2004.05.017}
  {\path{doi:10.1016/j.jcp.2004.05.017}}.

\bibitem{Ohwada2004Kobayashi}
T.~Ohwada, S.~Kobayashi, Management of discontinuous reconstruction in kinetic
  schemes, J. Comput. Phys. 197~(1) (2004) 116--138.
\newblock \href {https://doi.org/10.1016/j.jcp.2003.11.020}
  {\path{doi:10.1016/j.jcp.2003.11.020}}.

\bibitem{Ohwada2006}
T.~Ohwada, S.~Fukata, Simple derivation of high-resolution schemes for
  compressible flows by kinetic approach, J. Comput. Phys. 211~(2) (2006)
  424--447.
\newblock \href {https://doi.org/10.1016/j.jcp.2005.04.026}
  {\path{doi:10.1016/j.jcp.2005.04.026}}.

\bibitem{Frisch1986}
U.~Frisch, B.~Hasslacher, Y.~Pomeau, Lattice-gas automata for the navier-stokes
  equation, Phys. Rev. Lett. 56~(14) (1986) 1505.
\newblock \href {https://doi.org/10.1103/PhysRevLett.56.1505}
  {\path{doi:10.1103/PhysRevLett.56.1505}}.

\bibitem{Qian1992}
Y.~H. Qian, D.~d'Humi{\`e}res, P.~Lallemand, Lattice bgk models for
  navier-stokes equation, Europhys. Lett. 17~(6) (1992) 479--484.
\newblock \href {https://doi.org/10.1209/0295-5075/17/6/001}
  {\path{doi:10.1209/0295-5075/17/6/001}}.

\bibitem{Higuera1989}
F.~Higuera, S.~Succi, R.~Benzi, Lattice gas dynamics with enhanced collisions,
  Europhys. Lett. 9 (1989) 345--349.
\newblock \href {https://doi.org/10.1209/0295-5075/9/4/008}
  {\path{doi:10.1209/0295-5075/9/4/008}}.

\bibitem{Benzi1992}
R.~Benzi, S.~Succi, M.~Vergassola, The lattice boltzmann equation: theory and
  applications, Phys. Rep. 222 (1992) 145--197.
\newblock \href {https://doi.org/doi:10.1016/0370-1573(92)90090-m}
  {\path{doi:doi:10.1016/0370-1573(92)90090-m}}.

\bibitem{Succi2001}
S.~Succi, The lattice Boltzmann equation: for fluid dynamics and beyond, Oxford
  university press, 2001.

\bibitem{Shan2006}
X.~Shan, X.-F. Yuan, H.~Chen, Kinetic theory representation of hydrodynamics: a
  way beyond the navier--stokes equation, J. Fluid Mech. 550 (2006) 413--441.
\newblock \href {https://doi.org/10.1017/S0022112005008153}
  {\path{doi:10.1017/S0022112005008153}}.

\bibitem{Xu2018}
Y.~Gan, A.~Xu, G.~Zhang, Y.~Zhang, S.~Succi, Discrete boltzmann trans-scale
  modeling of high-speed compressible flows, Phys. Rev. E 95 (2018) 053312.
\newblock \href {https://doi.org/10.1103/PhysRevE.97.053312}
  {\path{doi:10.1103/PhysRevE.97.053312}}.

\bibitem{Xu2019}
Y.~Zhang, A.~Xu, G.~Zhang, Y.~Gan, Z.~Chen, S.~Succi, Entropy production in
  thermal phase separation: a kinetic-theory approach, Soft Matter 15 (2019)
  2245--2259.
\newblock \href {https://doi.org/10.1039/C8SM02637H}
  {\path{doi:10.1039/C8SM02637H}}.

\bibitem{Feuchter2016}
C.~Feuchter, W.~Schleifenbaum, High-order lattice boltzmann models for
  wall-bounded flows at finite knudsen numbers, Phys. Rev. E 94~(1) (2016)
  013304.
\newblock \href {https://doi.org/10.1103/PhysRevE.94.013304}
  {\path{doi:10.1103/PhysRevE.94.013304}}.

\bibitem{Ambrus2014}
V.~Ambrus, V.~Sofonea, Implementation of diffuse-reflection boundary conditions
  using lattice boltzmann models based on half-space gauss--laguerre
  quadratures, Phys. Rev. E 89 (2014) 041301(R).
\newblock \href {https://doi.org/10.1103/PhysRevE.96.013311}
  {\path{doi:10.1103/PhysRevE.96.013311}}.

\bibitem{Ambrus2016}
V.~Ambrus, V.~Sofonea, Lattice boltzmann models based on half-range
  gauss--hermite quadratures, J. Comp. Phys. 316 (2016) 760--768.
\newblock \href {https://doi.org/10.1016/j.jcp.2016.04.010}
  {\path{doi:10.1016/j.jcp.2016.04.010}}.

\bibitem{Ambrus2016mixed}
V.~E. Ambrus, V.~Sofonea, Application of mixed quadrature lattice boltzmann
  models for the simulation of poiseuille flow at non-negligible values of the
  knudsen number, J. Comp. Sci. 17 (2016) 403--417.
\newblock \href {https://doi.org/10.1016/j.jocs.2016.03.016}
  {\path{doi:10.1016/j.jocs.2016.03.016}}.

\bibitem{Mont2015}
A.~Montessori, P.~Prestininzi, M.~{La Rocca}, S.~Succi, Lattice boltzmann
  approach for complex nonequilibrium flows, Phys. Rev. E 92 (2015) 043308.
\newblock \href {https://doi.org/10.1103/PhysRevE.92.043308}
  {\path{doi:10.1103/PhysRevE.92.043308}}.

\bibitem{Rivet2001}
J.-P. Rivet, J.~P. Boon, Lattice gas hydrodynamics, Cambridge University Press,
  2001.

\bibitem{Broadwell1964shock}
J.~E. Broadwell, Shock structure in a simple discrete velocity gas, Phys.
  Fluids 7~(8) (1964) 1243--1247.
\newblock \href {https://doi.org/10.1063/1.1711368}
  {\path{doi:10.1063/1.1711368}}.

\bibitem{Gatignol1975}
R.~Gatignol, Th{\'e}orie cin{\'e}tique des gaz {\`a} r{\'e}partition
  discr{\`e}te de vitesses, Springer verlag, 1975.
\newblock \href {https://doi.org/10.1007/3-540-07156-3}
  {\path{doi:10.1007/3-540-07156-3}}.

\bibitem{Meng2011}
J.~Meng, Y.~Zhang, X.~Shan, Multiscale lattice boltzmann approach to modeling
  gas flows, Phys. Rev. E 83 (2011) 046701.
\newblock \href {https://doi.org/10.1103/PhysRevE.83.046701}
  {\path{doi:10.1103/PhysRevE.83.046701}}.

\bibitem{Staso2016long}
G.~Di~Staso, H.~J.~H. Clercx, S.~Succi, F.~Toschi, {DSMC--LBM} mapping scheme
  for rarefied and non-rarefied gas flows, J. Comp. Sci. 17 (2016) 357--369.
\newblock \href {https://doi.org/10.1016/j.jocs.2016.04.011}
  {\path{doi:10.1016/j.jocs.2016.04.011}}.

\bibitem{Staso2018}
G.~Di~Staso, S.~Srivastava, E.~Arlemark, H.~J.~H. Clercx, F.~Toschi, Hybrid
  lattice boltzmann-direct simulation monte carlo approach for flows in
  three-dimensional geometries, Comput. Fluids (2018).
\newblock \href {https://doi.org/10.1016/j.compfluid.2018.03.043}
  {\path{doi:10.1016/j.compfluid.2018.03.043}}.

\bibitem{Succi2016}
S.~Succi, Lattice boltzmann beyond navier-stokes: Where do we stand?, in: AIP
  Conference Proceedings, Vol. 1786, AIP Publishing, 2016, p. 030001.
\newblock \href {https://doi.org/10.1063/1.4967538}
  {\path{doi:10.1063/1.4967538}}.

\bibitem{Ilyin2018}
O.~Ilyin, A method for simulating the dynamics of rarefied gas based on lattice
  boltzmann equations and the bgk equation, Comp. Math. and Math. Phys. 58
  (2018) 1817--1827.
\newblock \href {https://doi.org/10.1134/S0965542518110052}
  {\path{doi:10.1134/S0965542518110052}}.

\bibitem{Aristov2019}
V.~V. Aristov, O.~V. Ilyin, O.~A. Rogozin, A hybrid numerical scheme based on
  coupling discrete-velocities models for the bgk and lbgk equations, in: AIP
  Conference Proceedings, Vol. 2132, AIP Publishing, 2019, p. 060007.
\newblock \href {https://doi.org/10.1063/1.5119547}
  {\path{doi:10.1063/1.5119547}}.

\bibitem{Succi1992}
F.~Nannelli, S.~Succi, The lattice boltzmann equation on irregular lattices, J.
  Stat. Phys. 68~(3-4) (1992) 401--407.
\newblock \href {https://doi.org/10.1007/BF01341755}
  {\path{doi:10.1007/BF01341755}}.

\bibitem{Peng1999}
G.~Peng, H.~Xi, C.~Duncan, S.-H. Chou, Finite volume scheme for the lattice
  boltzmann method on unstructured meshes, Phys. Rev. E 59~(4) (1999) 4675.
\newblock \href {https://doi.org/10.1103/PhysRevE.59.4675}
  {\path{doi:10.1103/PhysRevE.59.4675}}.

\bibitem{Patil2009}
D.~V. Patil, K.~Lakshmisha, Finite volume tvd formulation of lattice boltzmann
  simulation on unstructured mesh, J. Comput. Phys. 228~(14) (2009) 5262--5279.
\newblock \href {https://doi.org/10.1016/j.jcp.2009.04.008}
  {\path{doi:10.1016/j.jcp.2009.04.008}}.

\bibitem{Li2016}
W.~Li, L.-S. Luo, Finite volume lattice boltzmann method for nearly
  incompressible flows on arbitrary unstructured meshes, Commun. Comput. Phys.
  20~(2) (2016) 301--324.
\newblock \href {https://doi.org/10.4208/cicp.211015.040316a}
  {\path{doi:10.4208/cicp.211015.040316a}}.

\bibitem{Cercignani2000}
C.~Cercignani, Rarefied gas dynamics: from basic concepts to actual
  calculations, Cambridge University Press, 2000.

\bibitem{Sone2007}
Y.~Sone, Molecular gas dynamics: theory, techniques, and applications,
  Birkh\"{a}user, Boston, 2007.

\bibitem{Krook1954}
P.~L. Bhatnagar, E.~P. Gross, M.~Krook, A model for collision processes in
  gases. i. small amplitude processes in charged and neutral one-component
  systems, Phys. Rev. 94 (1954) 511--525.
\newblock \href {https://doi.org/10.1103/PhysRev.94.511}
  {\path{doi:10.1103/PhysRev.94.511}}.

\bibitem{Welander1954}
P.~Welander, On the temperature jump in a rarefied gas, Arkiv Fysik 7 (1954)
  507--553.

\bibitem{Cabannes1980}
H.~Cabannes, The discrete boltzmann equation (theory and applications), Lecture
  notes (1980).

\bibitem{Aristov2001}
V.~V. Aristov, Direct methods for solving the Boltzmann equation and study of
  nonequilibrium flows, Kluwer Academic Publishers, Dordrecht, 2001.

\bibitem{Mieussens2000}
L.~Mieussens, Discrete velocity model and implicit scheme for the bgk equation
  of rarefied gas dynamics, Math. Models Methods Appl. Sci. 10~(08) (2000)
  1121--1149.
\newblock \href {https://doi.org/10.1142/S0218202500000562}
  {\path{doi:10.1142/S0218202500000562}}.

\bibitem{Charrier1999}
P.~Charrier, B.~Dubroca, J.~L. Feugeas, Levermore's moment closure of discrete
  boltzmann equations for non-equilibrium kinetic flows, in: 21st international
  symposium on rarefied gas dynamics, Vol.~1, 1999, pp. 39--40.

\bibitem{Kruger2017}
T.~Kr{\"u}ger, H.~Kusumaatmaja, A.~Kuzmin, O.~Shardt, G.~Silva, E.~Viggen, The
  Lattice Boltzmann Method. Principles and Practice, Springer, 2017.

\bibitem{He1997}
X.~He, L.-S. Luo, A priori derivation of the lattice boltzmann equation, Phys.
  Rev. E 55~(6) (1997) R6333--R6336.
\newblock \href {https://doi.org/10.1103/PhysRevE.55.R6333}
  {\path{doi:10.1103/PhysRevE.55.R6333}}.

\bibitem{Shan1998}
X.~Shan, X.~He, Discretization of the velocity space in the solution of the
  boltzmann equation, Phys. Rev. Lett. 80 (1998) 65--68.
\newblock \href {https://doi.org/10.1103/PhysRevLett.80.65}
  {\path{doi:10.1103/PhysRevLett.80.65}}.

\bibitem{Shan2010}
X.~Shan, General solution of lattices for cartesian lattice
  bhatanagar--gross--krook models, Phys. Rev. E 81 (2010) 036702.
\newblock \href {https://doi.org/10.1103/PhysRevE.81.036702}
  {\path{doi:10.1103/PhysRevE.81.036702}}.

\bibitem{Karlin1999}
I.~Karlin, S.~Succi, On the {H-theorem} in lattice kinetic theory, Riv. Mat .
  Univ. Parma 6~(2) (1999) 143--154.

\bibitem{Chikatamarla2006}
S.~S. Chikatamarla, I.~V. Karlin, Entropy and galilean invariance of lattice
  boltzmann theories, Phys. Rev. Lett. 97 (2006) 190601.
\newblock \href {https://doi.org/10.1103/PhysRevLett.97.190601}
  {\path{doi:10.1103/PhysRevLett.97.190601}}.

\bibitem{Chikatamarla2009}
S.~S. Chikatamarla, I.~V. Karlin, Lattices for the lattice boltzmann method,
  Phys. Rev. E 79~(4) (2009) 046701.
\newblock \href {https://doi.org/10.1103/PhysRevE.79.046701}
  {\path{doi:10.1103/PhysRevE.79.046701}}.

\bibitem{Latt2006}
J.~Latt, B.~Chopard, Lattice boltzmann method with regularized pre-collision
  distribution functions, Math. Comp. Simul. 72 (2006) 165--168.
\newblock \href {https://doi.org/10.1016/j.matcom.2006.05.017}
  {\path{doi:10.1016/j.matcom.2006.05.017}}.

\bibitem{Chen2006}
H.~Chen, R.~Zhang, I.~Staroselsky, M.~Jhon, Recovery of full rotational
  invariance in lattice boltzmann formulations for high knudsen number flows,
  Phys. A 362 (2006) 125--131.
\newblock \href {https://doi.org/10.1016/j.physa.2005.09.008}
  {\path{doi:10.1016/j.physa.2005.09.008}}.

\bibitem{Zhang2006}
R.~Zhang, X.~Shan, H.~Chen, Efficient kinetic method for fluid simulation
  beyond the navier-stokes equation, Phys. Rev E 74 (2006) 046703.
\newblock \href {https://doi.org/10.1103/PhysRevE.74.046703}
  {\path{doi:10.1103/PhysRevE.74.046703}}.

\bibitem{Mattila2017}
K.~Mattila, P.~Philippi, L.~{Hegele Jr.}, High-order regularization in
  lattice-boltzmann equations, Phys. Fluids 29 (2017) 046103.
\newblock \href {https://doi.org/10.1063/1.4981227}
  {\path{doi:10.1063/1.4981227}}.

\bibitem{Bobylev2001}
A.~V. Bobylev, T.~Ohwada, The error of the splitting scheme for solving
  evolutionary equations, Appl. Math. Lett. 14~(1) (2001) 45--48.
\newblock \href {https://doi.org/10.1016/S0893-9659(00)00110-5}
  {\path{doi:10.1016/S0893-9659(00)00110-5}}.

\bibitem{LeVeque2002}
R.~J. LeVeque, et~al., Finite volume methods for hyperbolic problems, Vol.~31,
  Cambridge university press, 2002.

\bibitem{Baranger2019}
C.~Baranger, N.~H{\'e}rouard, J.~Mathiaud, L.~Mieussens, Numerical boundary
  conditions in finite volume and discontinuous galerkin schemes for the
  simulation of rarefied flows along solid boundaries, Math. Comput. Simul 159
  (2019) 136--153.
\newblock \href {https://doi.org/10.1016/j.matcom.2018.11.011}
  {\path{doi:10.1016/j.matcom.2018.11.011}}.

\bibitem{Inamuro1990}
T.~Inamuro, B.~Sturtevant, Numerical study of discrete-velocity gases, Phys.
  Fluids 2~(12) (1990) 2196--2203.
\newblock \href {https://doi.org/10.1063/1.857825}
  {\path{doi:10.1063/1.857825}}.

\bibitem{Aristov1980}
V.~V. Aristov, F.~G. Tcheremissine, Conservative splitting method for solving
  the boltzmann equation, USSR Comp. Math. Math. Phys. 20~(1) (1980) 208--225.
\newblock \href {https://doi.org/10.1016/0041-5553(80)90074-9}
  {\path{doi:10.1016/0041-5553(80)90074-9}}.

\bibitem{Kim2011}
C.~Kim, Formation and propagation of discontinuity for boltzmann equation in
  non-convex domains, Commun. Math. Phys. 308~(3) (2011) 641--701.
\newblock \href {https://doi.org/10.1007/s00220-011-1355-1}
  {\path{doi:10.1007/s00220-011-1355-1}}.

\bibitem{Guo2017}
Y.~Guo, C.~Kim, D.~Tonon, A.~Trescases, Regularity of the boltzmann equation in
  convex domains, Inventiones mathematicae 207~(1) (2017) 115--290.
\newblock \href {https://doi.org/10.1007/s00222-016-0670-8}
  {\path{doi:10.1007/s00222-016-0670-8}}.

\bibitem{Takata2016}
I.-K. Chen, H.~Funagane, T.-P. Liu, S.~Takata, Singularity of the velocity
  distribution function in molecular velocity space, Commun. Math. Phys.
  341~(1) (2016).
\newblock \href {https://doi.org/10.1007/s00220-015-2476-8}
  {\path{doi:10.1007/s00220-015-2476-8}}.

\bibitem{Takata2014}
I.-K. Chen, T.-P. Liu, S.~Takata, Boundary singularity for thermal
  transpiration problem of the linearized boltzmann equation., Archive for
  Rational Mechanics \& Analysis 212~(2) (2014).
\newblock \href {https://doi.org/10.1007/s00205-013-0714-9}
  {\path{doi:10.1007/s00205-013-0714-9}}.

\bibitem{Rogozin2019hybrid}
O.~Rogozin, {Hybrid finite-volume solver based on the lattice-Boltzmann and
  discrete-velocity methods for the Couette problem} (2019).
\newblock \href {https://doi.org/10.5281/zenodo.3471926}
  {\path{doi:10.5281/zenodo.3471926}}.

\bibitem{Luo2015}
W.~Li, L.-S. Luo, J.~Shen, Accurate solution and approximations of the
  linearized bgk equation for steady couette flow, Comput. Fluids 111 (2015)
  18--32.
\newblock \href {https://doi.org/10.1016/j.compfluid.2014.12.018}
  {\path{doi:10.1016/j.compfluid.2014.12.018}}.

\bibitem{Luo2016}
S.~Jiang, L.-S. Luo, Analysis and accurate numerical solutions of the integral
  equation derived from the linearized bgkw equation for the steady couette
  flow, J. Comput. Phys. 316 (2016) 416--434.
\newblock \href {https://doi.org/10.1016/j.jcp.2016.04.011}
  {\path{doi:10.1016/j.jcp.2016.04.011}}.

\bibitem{Rogozin2019couette}
O.~Rogozin, {Solver for the linear Couette-flow problem for a rarefied BGK gas}
  (2019).
\newblock \href {https://doi.org/10.5281/zenodo.3471914}
  {\path{doi:10.5281/zenodo.3471914}}.

\bibitem{Ohwada1990}
Y.~Sone, S.~Takata, T.~Ohwada, Numerical analysis of the plane couette flow of
  a rarefied gas on the basis of the linearized boltzmann equation for
  hard-sphere molecules, Eur. J. Mech. B/Fluids 9 (1990) 273--288.

\bibitem{Wu2014}
L.~Wu, J.~M. Reese, Y.~Zhang, Solving the boltzmann equation deterministically
  by the fast spectral method: application to gas microflows, J. Fluid Mech.
  746 (2014) 53--84.
\newblock \href {https://doi.org/10.1017/jfm.2014.79}
  {\path{doi:10.1017/jfm.2014.79}}.

\bibitem{Rogozin2016}
O.~Rogozin, Numerical analysis of the nonlinear plane couette-flow problem of a
  rarefied gas for hard-sphere molecules, Eur. J. Mech. B/Fluids 60 (2016)
  148--163.
\newblock \href {https://doi.org/10.1016/j.euromechflu.2016.06.011}
  {\path{doi:10.1016/j.euromechflu.2016.06.011}}.

\bibitem{Sharipov2013}
F.~Sharipov, J.~L. Strapasson, Benchmark problems for mixtures of rarefied
  gases. i. couette flow, Phys. Fluids 25~(2) (2013) 027101.
\newblock \href {https://doi.org/10.1017/jfm.2014.79}
  {\path{doi:10.1017/jfm.2014.79}}.

\bibitem{Su2018}
W.~Su, P.~Wang, H.~Liu, L.~Wu, Accurate and efficient computation of the
  boltzmann equation for couette flow: influence of intermolecular potentials
  on knudsen layer function and viscous slip coefficient, J. Comput. Phys.
  (2018).
\newblock \href {https://doi.org/10.1016/j.jcp.2018.11.015}
  {\path{doi:10.1016/j.jcp.2018.11.015}}.

\bibitem{Zhang2014}
J.~Meng, N.~Dongari, J.~M. Reese, Y.~Zhang, Breakdown parameter for kinetic
  modeling of multiscale gas flows, Phys. Rev. E 89~(6) (2014) 063305.
\newblock \href {https://doi.org/10.1103/PhysRevE.89.063305}
  {\path{doi:10.1103/PhysRevE.89.063305}}.

\bibitem{Chen2010}
F.~Chen, A.~Xu, G.~Zhang, Y.~Li, S.~Succi, Multiple-relaxation-time lattice
  boltzmann approach to compressible flows with flexible specific-heat ratio
  and prandtl number, Europhys. Lett. 90 (2010) 54003.
\newblock \href {https://doi.org/10.1209/0295-5075/90/54003}
  {\path{doi:10.1209/0295-5075/90/54003}}.

\bibitem{Frapolli2015}
N.~Frapolli, S.~Chikatamarla, I.~Karlin, Entropic lattice boltzmann model for
  compressible flows, Phys. Rev. E 92 (2015) 061301(R).
\newblock \href {https://doi.org/10.1103/PhysRevE.92.061301}
  {\path{doi:10.1103/PhysRevE.92.061301}}.

\bibitem{Frapolli2016}
N.~Frapolli, S.~Chikatamarla, I.~Karlin, Entropic lattice boltzmann model for
  gas dynamics: Theory, boundary conditions, and implementation, Phys. Rev. E
  93 (2016) 063302.
\newblock \href {https://doi.org/10.1103/PhysRevE.93.063302}
  {\path{doi:10.1103/PhysRevE.93.063302}}.

\bibitem{Aristov1977}
V.~V. Aristov, Method of adaptative meshes in velocity space for the intense
  shock wave problem, USSR Comput. Math. Math. Phys. 17~(4) (1977) 261--267.

\bibitem{Kolobov2013}
R.~R. Arslanbekov, V.~I. Kolobov, A.~A. Frolova, Kinetic solvers with adaptive
  mesh in phase space, Phys. Rev. E 88~(6) (2013) 063301.
\newblock \href {https://doi.org/10.1103/PhysRevE.88.063301}
  {\path{doi:10.1103/PhysRevE.88.063301}}.

\bibitem{Baranger2014}
C.~Baranger, J.~Claudel, N.~H{\'e}rouard, L.~Mieussens, Locally refined
  discrete velocity grids for stationary rarefied flow simulations, J. Comput.
  Phys. 257 (2014) 572--593.
\newblock \href {https://doi.org/10.1016/j.jcp.2013.10.014}
  {\path{doi:10.1016/j.jcp.2013.10.014}}.

\bibitem{Willis1962}
D.~R. Willis, Comparison of kinetic theory analyses of linearized couette flow,
  Phys. Fluids 5~(2) (1962) 127--135.
\newblock \href {https://doi.org/10.1063/1.1706585}
  {\path{doi:10.1063/1.1706585}}.

\bibitem{Abramowitz1972}
M.~Abramowitz, I.~A. Stegun, Handbook of mathematical functions: with formulas,
  graphs, and mathematical tables, National Bureau of Standards, 1972.

\end{thebibliography}

\end{document}